\documentclass[fleqn,usenatbib]{mnras}

\usepackage[T1]{fontenc}
\usepackage{ae,aecompl}

\usepackage{graphicx}	
\usepackage{caption}
\usepackage{subcaption}
\usepackage{amsmath}	
\usepackage{amssymb}	
\usepackage{float}
\usepackage{ftnxtra} 
\usepackage{fnpos} 
\usepackage{textcomp} 

\usepackage{ctable}

\newcommand{\disc}{\ensuremath{\textrm{disc}}}
\newcommand{\tot}{\ensuremath{\textrm{tot}}}
\newcommand{\BB}{\ensuremath{\textrm{BB}}}
\newcommand{\Boltz}{\ensuremath{\textrm{B}}}
\newcommand{\fracd}{\ensuremath{\textrm{d}}}

\newcommand{\const}{\ensuremath{\textrm{const}}}
\newcommand{\fracref}{\ensuremath{\textrm{ref}}}
\newcommand{\dust}{\ensuremath{\textrm{dust}}}
\newcommand{\fraclog}{\ensuremath{\textrm{log}}}
\newcommand{\natlog}{\ensuremath{\textrm{ln}}}

\newcommand{\mum}{\,\mu\hbox{m}}

\newcommand{\AU}{\,\hbox{AU}}

\newcommand{\AAp}      {Astron. Astrophys.}

\newcommand{\AJ}       {Astron. J.}

\newcommand{\ApJ}      {Astrophys. J.}
\newcommand{\ApJL}      {Astrophys. J. Lett.}

\newcommand{\ARAA}     {Ann. Rev. Astron. Astrophys.}

\newcommand{\MNRAS}    {MNRAS}

\title[Observability of Debris Discs around M-stars]{Observability of Dusty Debris Discs around M-stars}

\author[P. Luppe et al.]{Patricia Luppe,$^{1}$\thanks{E-mail: patricia.luppe@posteo.net}
Alexander V. Krivov,$^{1}$
Mark Booth$^{1}$
and \newauthor Jean-Fran\c{c}ois Lestrade$^{2}$
\\
$^{1}$Astrophysikalisches
Institut und Universit\"atssternwarte, Friedrich-Schiller-Universit\"at Jena, Schillerg\"a\ss{}chen 2-3, D-07745 Jena, \\Germany\\
$^{2}$Observatoire de Paris, CNRS, 61 Av. de l'Observatoire, 75014 Paris, France\\
}

\date{Accepted XXX. Received YYY; in original form \today}

\pubyear{XXX}

\begin{document}
\label{firstpage}
\pagerange{\pageref{firstpage}--\pageref{lastpage}}
\maketitle

\begin{abstract}
Debris discs are second generation dusty discs formed by collisions of planetesimals. Many debris discs have been found and resolved around hot and solar-type stars. However, only a handful have been discovered around M-stars, and the reasons for their paucity remain unclear.
Here we check whether the sensitivity and wavelength coverage of present-day telescopes are simply unfavourable for detection of these discs or if they are truly rare. We approach this question by looking at the \emph{Herschel}/DEBRIS survey that has searched for debris discs including M-type stars. Assuming that these cool-star discs are ``similar'' to those of the hotter stars in some sense (i.e., in terms of dust location, temperature, fractional luminosity, or mass), we check whether this survey should have found them.
With our procedure we can reproduce the $2.1^{+4.5}_{-1.7}$\% detection rate of M-star debris discs of the DEBRIS survey, which implies that these discs can indeed be similar to discs around hotter stars and just avoid detection. We then apply this procedure to IRAM NIKA-2 and ALMA bands~3, 6 and 7 to predict possible detection rates and give recommendations for future observations. We do not favour observing with IRAM, since it leads to detection rates lower than for the DEBRIS survey, with 0.6\%--4.5\% for a 15~min observation. ALMA observations, with detection rates 0.9\%--7.2\%,  do not offer a significant improvement either, and so we conclude that more sensitive far-infrared and single dish sub-millimetre telescopes are necessary to discover the missing population of M-star debris discs.
\end{abstract}




\section{Introduction}

In many respects, M-star planetary systems differ from systems around earlier-type stars. They seem to contain different populations of planets, as many M-stars host Earth- to Neptune-mass planets but only a few harbor gas giants \citep{bonfils-et-al-2013,mulders-et-al-2015,dressing-charbonneau-2015,winn-2018}. Another peculiarity is related to debris discs around these cool stars. Debris discs are second generation, dusty circumstellar discs formed by collisions of planetesimals that have formed in previously existing planet-forming discs. While during the last few decades many debris discs have been found and resolved around A to K-type stars \citep{matthews-et-al-2014,hughes-et-al-2018}, only a handful of them have been discovered around M-stars.

We start with a brief summary of previous searches of debris discs around M-stars. Many of these surveys, such as \citeauthor{plavchan-et-al-2005} (\citeyear{plavchan-et-al-2005},  \citeyear{plavchan-et-al-2009}), \citet{lestrade-et-al-2009} and \citet{avenhaus-et-al-2012}, did not result in detections. Other surveys have been more successful. \citet{forbrich-et-al-2008} detected 11 M-star debris discs with Spitzer, including 9 new disc candidates. \citet{theissen-west-2014} used data released by the Wide-field Infrared Survey Explorer (WISE) to look for excesses around field M-stars. In their sample of 70841 M-stars they detected excesses around 175 stars. \citet{binks-jeffries-2017} also used WISE data and found 7 debris disc candidates in nearby young moving groups. More recently, \citet{zuckerman-et-al-2019} used WISE data to demonstrate infrared excesses around 9 stars in the $\chi^1$ cluster, many around stars that are in binary systems.\citet{nibauer-et-al-2020} used Planck data to statistically determine that $\sim$10\% of A-M stars within 80~pc of the Sun (of which almost a half are M stars) have detectable debris discs. Their method cannot conclusively pinpoint which stars have debris discs, but they do propose a number of specific M stars as likely candidates. Three surveys searched for M-star debris discs with \emph{Herschel}/PACS. The DEBRIS survey, which we use in this paper, probed 94 M-stars at 100 and 160$\mu$m (Lestrade et al., in prep.). It yielded the detection of a disc around GJ 581 \citep{lestrade-et-al-2012} and Fomalhaut C \citep{kennedy-et-al-2014}. Another PACS survey presented by \citet{kennedy-et-al-2018b} examined 21 M-stars with known radial-velocity planets and yielded two disc candidates around GJ 433 and GJ 649. A PACS survey by \citet{tanner-et-al-2020} observed 20 stars between spectral types K5 and M5 and yielded three disc candidates around CP-72 2713\footnote{Although there is debate over whether this has a spectral type of K7 \citep{torres-et-al-2006,pecaut-mamajek-2013} or M0 \citep{gaidos-et-al-2014}.\label{fnspt}}, GJ 784 and GJ 707. In the sub-mm, 32 M-stars were observed at 850$\mu$m with SCUBA at JCMT and at 1.2~mm with MAMBO at the IRAM telescope \citep{lestrade-et-al-2006}, leading to one disc candidate around GJ 842.2. At present,  there are still only five debris disc detections around M-stars based on more than one independent observation: AU Mic (\citeauthor{kalas-et-al-2004} \citeyear{kalas-et-al-2004}, \citeauthor{liu-et-al-2004} \citeyear{liu-et-al-2004}, \citeauthor{fitzgerald-et-al-2007b} \citeyear{fitzgerald-et-al-2007b}, \citeauthor{macgregor-et-al-2013} \citeyear{macgregor-et-al-2013}), GJ 581 \citep{lestrade-et-al-2012,kennedy-et-al-2018b}, Fomalhaut~C (\citeauthor{kennedy-et-al-2014} \citeyear{kennedy-et-al-2014}; Cronin-Coltsmann et al., in prep.), TWA 7 \citep{matthews-et-al-2007a,bayo-et-al-2019,matra-et-al-2019} and CP-72 2713\footnotemark[1] \citep{moor-et-al-2020,tanner-et-al-2020}. Apart from that, several other detections \citep[such as TWA~25;][]{choquet-et-al-2016} were claimed, but these are pending confirmation. 

The reasons for the paucity of debris disc detection around M-stars remain unclear. 
It might be possible that M-star debris discs are no less common than those around earlier-type hosts, and that the lack of detections is solely caused by observational limitations.  \citet{morey-lestrade-2014} investigated this possibility with detailed modelling. They assumed a debris disc created by a planetesimal belt that is undergoing a steady-state collisional cascade and employed the analytic model of \citet{wyatt-et-al-2007} for the disc collisional evolution. By using distributions of mean disc radii and initial disc masses they generated a synthetic disc population. They fit their model to the fractional luminosities of 34 A-star debris disc from \citet{su-et-al-2006} and 28 FGK-star debris discs from \citet{trilling-et-al-2008}. All of these stars were observed with Spitzer at 24 and 70$\mu$m. With the assumption that the same disc population that was found around AFGK-stars is located around M-stars they used the best fit parameters of their models to simulate total disc masses and mean disc radii of potential debris discs around the M-stars listed in \citet{gautier-et-al-2007} and \citet{lestrade-et-al-2006,lestrade-et-al-2009}. In this way they estimated the flux density of these potential discs at 70$\mu$m. They concluded that a significant population of M-star discs could exist and would still be consistent with the non-detections in preceding surveys.

Like \citet{morey-lestrade-2014}, here we check whether the sensitivity and wavelength coverage of present-day telescopes are simply unfavourable for detection of these discs or if they are truly rare. However, our approach is different to theirs and can be referred to as more empirical.
The idea is to investigate whether a true population of debris 
discs around M-stars could be ``similar'' in one or another sense to a population of debris discs around stars of earlier spectral types (AFGK-stars), and whether the paucity of M-star discs could be solely due to their limited detectability with the instruments used previously to search for them.
That ``similarity'' can be formulated in terms of dust location, temperature, fractional 
luminosity, or mass.
We formulate several hypotheses, with the simplest one being a hypothesis that discs around 
stars of all spectral types have comparable radii and dust masses.
Alternatively, we can assume some scalings of disc radii and/or masses with the spectral type, 
for instance take an expectation that discs of later-type stars are smaller 
and/or lighter than those around earlier-type primaries.
For each hypothesis, we generate an expected population of M-star debris discs, check their detectability with existing instruments, and compare the predicted detection rates with actual ones reported previously.
In this way, we select those hypotheses that are consistent with the previous surveys and reject the others that are not.
Finally, we use the preferred hypotheses to make testable predictions by calculating
the expected detection probability of M-star discs in future surveys with other instruments 
(ALMA and IRAM instead of \emph{Herschel}).

Section~2 describes important equations, our procedure, and the different hypotheses we make.
Section~3 presents the results for our tests with the DEBRIS survey and our predictions for observations with IRAM and ALMA.
Section~4 contains a discussion and section~5 lists our conclusions.

\section{Methods}
\label{chap:methods}

\subsection{Basic equations}
Two basic quantities characterizing a debris disc are dust temperature $T_{\fracd}$ and fractional luminosity $f_{\fracd}$. Both can be inferred from the observed spectral energy distributions (SEDs) by means of a straightforward SED modelling.
The fractional luminosity is related to the total cross section of the dust material $\sigma_{\tot}$ and the disc radius $R_\fracd$:
\begin{equation}
    f_{\fracd}=\frac{\sigma_{\tot}}{4 \pi R_{\fracd}^{2}}.
	\label{eq:frac_lum}
\end{equation}
In turn, $\sigma_{\tot}$ is directly proportional to the dust mass $M_{\fracd}$:
\begin{equation}
 M_{\fracd} \propto \sigma_{\tot},
\end{equation}
assuming that the dust size distribution is the same in all discs considered.
The dust temperature is determined by stellar luminosity $L_*$ and blackbody disc radius $R_{\BB}$:
\begin{equation}
    T_{\fracd}= \left(\frac{L_{*}}{16 \pi \sigma_{\Boltz}}\right)^{1/4} \left(R_{\BB}\right)^{-1/2},
	\label{eq:BB-temp}
\end{equation}
with $\sigma_{\Boltz}$ being the Stefan-Boltzmann constant.

The blackbody radius is the one the disc of temperature $T_{\fracd}$ would have if the grains were absorbing and emitting as blackbodies. However, this is not the case since any disc also includes dust particles that are smaller than the typical wavelengths of their thermal emission. Such grains are efficient absorbers of the incoming stellar radiation, but inefficient emitters of thermal radiation \citep[e.g.,][]{krivov-et-al-2008}. This makes these grains warmer than the blackbody approximation predicts, implying that the actual radius of a disc of temperature $T_{\fracd}$, which we denote as $R_{\fracd}$, must be larger than $R_{\BB}$ \citep{rodriguez-zuckerman-2012,booth-et-al-2013,pawellek-et-al-2014,pawellek-krivov-2015}. In this paper, for the conversion of $R_{\BB}$ to $R_{\fracd}$ we use the relation \citep[see Fig.~4b in][]{pawellek-et-al-2014}
\begin{equation}
    R_{\fracd}= \Gamma\, R_{\BB},
\qquad \mathrm{with} \qquad
  \Gamma = 2.0
         \left( L_{*} / L_{\odot} \right)^{-0.16}.
	\label{eq:Gamma}
\end{equation}
As shown by \citet{pawellek-et-al-2014}, this relation holds over a wide range of stellar luminosities, albeit with a large scatter. It is also approximately valid for two prominent debris discs around M-stars, AU Mic \citep{matthews-et-al-2015} and GJ~581 \citep{lestrade-et-al-2012}. We thus assume that Eq.~(\ref{eq:Gamma}) is applicable to all M-star discs considered in our analysis. Of course, this is just an assumption that we need to make to be able to estimate the temperature of the as yet undiscovered discs around M-stars from their expected ``physical'' radii.

To calculate the detection limits of the telescopes, it is sufficient to assume that dust emits as a blackbody. Denoting by $F_{\nu,\mathrm{min}}$ the minimum specific thermal emission flux detectable with a certain instrument at a frequency $\nu$, it is easy to express the minimum fractional luminosity of a disc that this instrument is able to detect:
\begin{equation}
    f_{\fracd, \mathrm{min}}= \frac{2 d^{2} c^{2}}{h \nu^{3}} 
    				\frac{\sigma_{\Boltz} T_{\fracd}^{4}}{L_{*}}
                                F_{\nu,\mathrm{min}}
    				\left( \exp \left[ \frac{h \nu}{k T_{\fracd}} \right] - 1 \right), 
	\label{eq:detection-limit}
\end{equation}
where $d$ is the distance to the star, $c$ the speed of light, and $h$ the Planck constant.

\subsection{Hypotheses}

We now formulate a set of hypotheses of what the population of M-star debris discs may look like.
To this end, we consider possible trends of disc radius and mass with the spectral type of disc host stars.
In doing so, we include both the relations reported for debris discs of AFGK-stars and those found in millimetre surveys of protoplanetary discs, which are progenitors to the debris discs in question.

We start with the disc radius, $R_{\fracd}$. It is still debated if a significant correlation between the spectral type and the debris disc radius exists. Taking ALMA- and SMA-resolved debris discs, \citeauthor{matra-et-al-2018} (\citeyear{matra-et-al-2018}, their Fig.~1) find a slight correlation between debris disc radius and stellar luminosity.
Others, such as \citeauthor{pawellek-et-al-2014} (\citeyear{pawellek-et-al-2014}, based on \emph{Herschel}-resolved debris discs, see their Fig. 2) and \citeauthor{hughes-et-al-2018} (\citeyear{hughes-et-al-2018}, their Fig.~4) do not find any correlation.
Therefore we state for hypotheses~1 to~5 that the disc radii of debris discs around M-stars correspond to those around the reference stars: $R_{\fracd}= \const$. If $R_{\BB}$ of the reference star is 100~AU, this leads to $R_{\fracd}^{M}=R_{\fracd}^{\fracref}= \Gamma\,R_{BB}^{\fracref}=\Gamma\cdot 100~\AU$. For the hypotheses~6 to~10 we use the radius-luminosity relation proposed by \citet{matra-et-al-2018}:
\begin{equation}
    R_{\fracd} \propto L_{*}^{0.19}.
	\label{eq:matra-R-L}
\end{equation}

\begin{table}
	\centering
	\caption{Summary of hypotheses. The second and the third column list the assumed scalings of the disc radius and dust cross section with the stellar mass. The fourth and the fifth column show how the dust temperature and fractional luminosity change from AFGK-star debris discs to M-star discs assuming these scalings.}
	\label{tab:summary_table}
	\begin{tabular}{lcccl} 
		\hline
		Hypothesis & $R_{\fracd}$ & $\sigma_{\tot}$ & $T_{\fracd}$ & $f_{\fracd}$ \\
		\hline
		 1 & const 					& const 					& decrease	& const \\
		 2 & const					& $\propto M_{*}$ 		& decrease	& decrease \\
		 3 & const 					& $\propto M_{*}^{1.3}$ 	& decrease	& decrease\\		
		 4 & const 					& $\propto M_{*}^{1.9}$ 	& decrease	& decrease \\
		 5 & const					& $\propto M_{*}^{2.7}$ 	& decrease	& decrease \\
		 6 & $\propto L_{*}^{0.19}$	& const 					& decrease	& increase \\
		 7 & $\propto L_{*}^{0.19}$	& $\propto M_{*}$		& decrease	& increase \\
		 8 & $\propto L_{*}^{0.19}$	& $\propto M_{*}^{1.3}$	& decrease	& slight in-/decrease\\
		 9 & $\propto L_{*}^{0.19}$	& $\propto M_{*}^{1.9}$	& decrease	& slight in-/decrease \\
		10 & $\propto L_{*}^{0.19}$	& $\propto M_{*}^{2.7}$	& decrease	& decrease \\
		\hline
	\end{tabular}
\end{table}

As far as the scaling of the disc mass is concerned, we used five different assumptions. For hypotheses~1 and~6, we assumed the mass of the M-star discs to be the same as for the reference stars. For hypotheses~2 and~7, the relation of \citet{williams-cieza-2011} for protoplanetary discs ($M_{\fracd} \propto M_{*}$), deduced from the \hbox{(sub-)mm} surveys between 2000 and 2009, was adopted. For the other hypotheses, we took relations listed in \citet{pascucci-et-al-2016}. They presented an ALMA observation of the 2~Myr-old Chamaeleon I star-forming region and found the relation $M_{\dust} \propto M_{*}^{1.3-1.9}$. Re-analysing all data of nearby star-forming regions that were available in the mm-range, they found that the 1-3~Myr-old Taurus, Lupus and Chamaeleon I regions all show similar $M_{\dust}-M_{*}$ dependence. We therefore used the slope of 1.3 for hypotheses~3 and~8 and the slope of 1.9 for hypotheses~4 and~9. A much higher slope, with $M_{\dust} \propto M_{*}^{2.7}$, was found by \citet{pascucci-et-al-2016} for the 10~Myr old Upper Sco association. We used this slope for our hypotheses~5 and 10. An overview of all hypotheses (i.e., combinations of the parameter scalings) is given in Table~\ref{tab:summary_table}.

\begin{figure*}
     \begin{center}
       \label{fig:third}
       \includegraphics[width=0.7\textwidth]{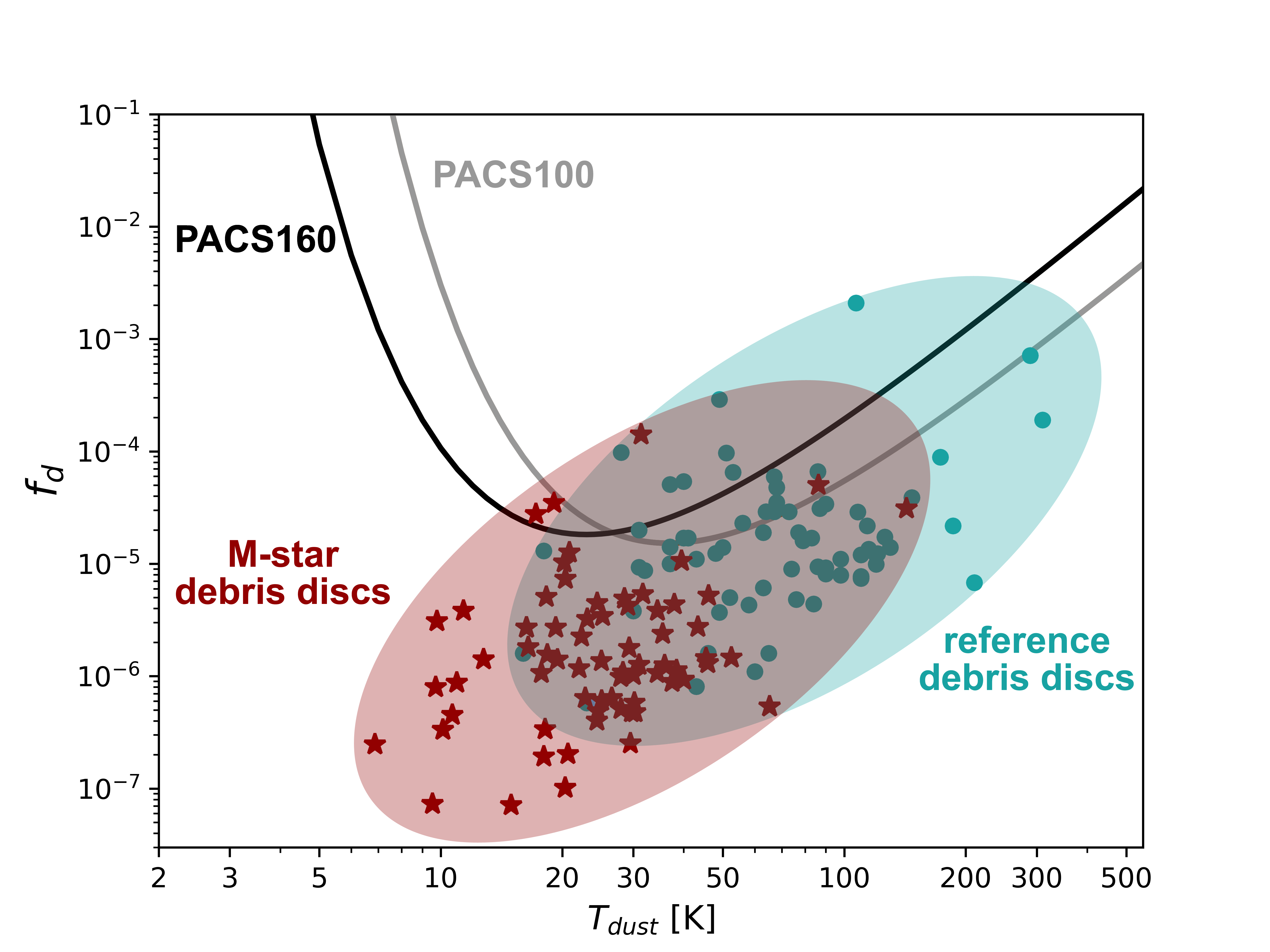}
    \end{center}
    \caption{Schematic that illustrates how we generate a population of M-star debris discs, which is expected in the framework of a certain hypothesis. This is the plot for a 15~min observation of GJ~447, an M4 star, at a distance of 3.37~pc, and for hypothesis~3. The brown U-shaped curve gives the limit for the \emph{Herschel}/PACS 100$\mu $m band, the black curve for the PACS 160$\mu $m band. The blue points represent the reference debris discs of the AFGK-stars found by DEBRIS and the pink stars are the calculated M-star debris discs. The shaded regions highlight the parameter space of the reference discs and the simulated discs in order to better visualise how this changes based on the assumed hypothesis. Observable are discs that are located above the detectability curves. The reference discs do not necessarily lie above these curves, because the curves were calculated for an M-star, GJ 447. Detectability curves change with $L_{*}$ and distance and have to be calculated independently for every star.}
   \label{fig:schematic}
\end{figure*}

\subsection{Procedure}

To check a specific hypothesis, we proceed as follows (Fig.~\ref{fig:schematic}).

1. First, we consider a sample of reference stars of earlier spectral types (we take AFGK-stars) with debris disc detections.
All their discs have known dust fractional luminosities $f^\text{ref}_\text{d}$
and temperatures~$T^\text{ref}_\text{d}$.

2. Then we take one of the observational programs that searched for debris discs around M-stars
(numbered 1, 2, ... $N$)
with a certain instrument and reported both detections and non-detections.
For every M-star observed, we know the stellar parameters.
We then take $f^\text{ref}_d$ and $T^\text{ref}_\text{d}$ and use Eqs.~(\ref{eq:frac_lum})--(\ref{eq:Gamma}) together with the scalings of Table~\ref{tab:summary_table}
to calculate $f^\text{M}_\text{d}$ and $T^\text{M}_\text{d}$,
i.e. the fractional luminosity and temperature that dust in the (scaled) reference discs would have if their central star were the M-star considered.
This yields a population of debris discs around an M-star that we expect in the framework of the selected hypothesis. 
We also know the specifications of the instrument 
used in the M-star survey, and the integration time.
This allows us to compute the detection limit (Eq.~\ref{eq:detection-limit}), which will be a 
U-shaped curve in the temperature -- fractional luminosity plane (Fig.~\ref{fig:schematic}). For all following calculations we used $3\sigma$ detection limits.
Note that the detection limit depends on the distance and luminosity of the M-star and so, the detectability 
curves will be different for different M-stars in the survey.

3. Finally, we compare the position of the generated cloud of fiducual M-star discs with that 
detectability curve and count the fraction of discs that would be observable, $p_i$
(where $i$ is a number of the M-star in the M-star survey).

4. The entire procedure is executed for all M-stars in the survey, yielding the detection
probabilities $p_i$ ($i=1$, $2$, ... $N$) and the average detection probability
$$\bar{p}^\text{M} = \frac{1}{N} \sum\limits_{i=1}^N p_i .$$

5. Finally, we multiply $\bar{p}^\text{M}$ with the detection rate of debris discs around 
reference stars, $p^\text{ref}$, to get the detection rate of discs around M-stars that we 
predict with a given hypothesis:
$$p^\text{M}_\text{pred} = \bar{p}^\text{M} p^\text{ref}.$$
This should be compared with the actual detection fraction of discs in the M-star survey, 
$p^\text{M}_\text{obs}$.
The closer 
$p^\text{M}_\text{pred}$
and
$p^\text{M}_\text{obs}$,
the more viable the hypothesis we made about the true population of discs around M-stars.


\section{Results}

In this section, we describe our sample of reference stars and the M-star debris disc survey.
We then apply the algorithm described above to test the hypotheses.
Finally, we make predictions for future IRAM and ALMA surveys of M-star debris discs.

\subsection{The DEBRIS sample}

The DEBRIS Open Time Key Program \citep{matthews-et-al-2010} was an unbiased \emph{Herschel} survey that searched for debris discs around A-, F-, G-, K- and M-stars (Table~\ref{tab:DEBRIS-sample+value-sources}). In this survey 83 A-stars \citep{thureau-et-al-2014}, 92 F-stars, 90 G-stars, 91 K-stars \citep{sibthorpe-et-al-2018} and 94 M-stars (Lestrade et al., in prep.) were observed at $100\mum$ and $160\mum$. In the originally planned unbiased sample 86 A-stars were listed \citep{phillips-et-al-2010}. Three of those stars were excluded from the survey because they were guaranteed time \emph{Herschel} targets: Vega \citep{sibthorpe-et-al-2010}, Fomalhaut \citep{acke-et-al-2012} and $\beta$ Pic \citep{vandenbussche-et-al-2010}. They nevertheless were included in all DEBRIS statistics, because leaving them out would have biased the sample. The same applies to the GK-stars. In the original unbiased sample 91 G-stars and 92 K-stars were listed. $\varepsilon$~Eridani \citep{greaves-et-al-2014b} and $\tau$~Ceti \citep{lawler-et-al-2014} were part of the \emph{Herschel} guaranteed time program. Both were included in all DEBRIS statistics for the same reasons as for the A-stars. Therefore we use all 86 A-stars, 91 G-stars and 92 K-stars for our calculations and statistics, too.
The number of observed stars and disc detections that we used for our analysis are listed in Table~\ref{tab:DEBRIS-sample+value-sources}. Five A-stars were identified to possess a two-component disc. For those stars we only included the cold disc component (\citeauthor{thureau-et-al-2014} \citeyear{thureau-et-al-2014}, their Table~4) in our calculations.
For three of the FGK-star discs a two-component fit was necessary to reproduce the observed excesses \citep{sibthorpe-et-al-2018}. In these cases  \citet{sibthorpe-et-al-2018} used the cooler of the two components for their analysis.


\begin{table*}
	\centering
	\caption{DEBRIS sample by spectral classes. In the left part of the table the DEBRIS sample is listed. For five A-stars reported as having two-component discs we only used the cold component for our calculation. The right part lists the sources we used for the stellar parameters.}
	\label{tab:DEBRIS-sample+value-sources}
	\begin{tabular}{lcccc||ccccccc}
		\hline
		stellar type & stars & detections & det. rate & refs & SpT & $L_{*}$ & $M_{*}$ & distance & $T_{\disc}$ & $f_{\fracd}$ & $R_{\BB}$ \\
		\hline
		A & 86 & 21 & $24.4\%^{+9.8}_{-8.1}$ & 1 & 1 & 6,7,8,9,10,11 & calc. from $L_{*}$ & - & 1 & 1 & 1\\
		F & 92 & 22 & $23.9\%^{+9.4}_{-7.8}$ & 2 & 2 & 6,13 & 13 or calc. from $L_{*}$ & - & 2 & 2 & 2\\
		G & 91 & 13 & $14.3\%^{+8.3}_{-6.0}$ & 2 & 2 & 6,13 & 13 or calc. from $L_{*}$ & - & 2 & 2 & 2 \\	
		K & 92 & 12 & $13.0\%^{+8.0}_{-5.7}$ & 2 & 2 & 6,13 & 13 or calc. from $L_{*}$ & 6,12 & 2 & 2 & 2\\
		M & 94 & 2 & $2.1\%^{+4.5}_{-1.7}$ & 2,3,4,5 & 3 & 13 & 13 & 6 & - & - & - \\
		\hline
	\end{tabular}
		\\[3mm]
\textbf{References:}
1:~\citet{thureau-et-al-2014},
2:~\citet{sibthorpe-et-al-2018},
3:~Lestrade et al. (in prep.), 
4:~\citet{kennedy-et-al-2014},
5:~\citet{lestrade-et-al-2012},
6:~\citet{gaia-2018},
7:~\citet{kennedy-et-al-2012b},
8:~\citet{difolco-et-al-2004},
9:~\citet{anderson-francis-2012},
10:~\citet{zorec-et-al-2012}, 
11:~\citep{boyajian-et-al-2012a},
12:~\citet{vanleeuwen-2007},
13:~Mamajek (Version 2018.12.10)\footnotemark[1]
\\
\textbf{Note:} 
The errors for the detection rates were calculated with the 95\% confidence limit, Jeffreys interval\footnotemark[2]
\end{table*}
\footnotetext[1]{\url{http://www.pas.rochester.edu/~emamajek/EEM_dwarf_UBVIJHK_colors_Teff.txt}}

\subsection{Tests of hypotheses with AFGK-stars}

\begin{figure}
     \begin{center}
            \hspace*{-5mm}\includegraphics[width=0.55\textwidth]{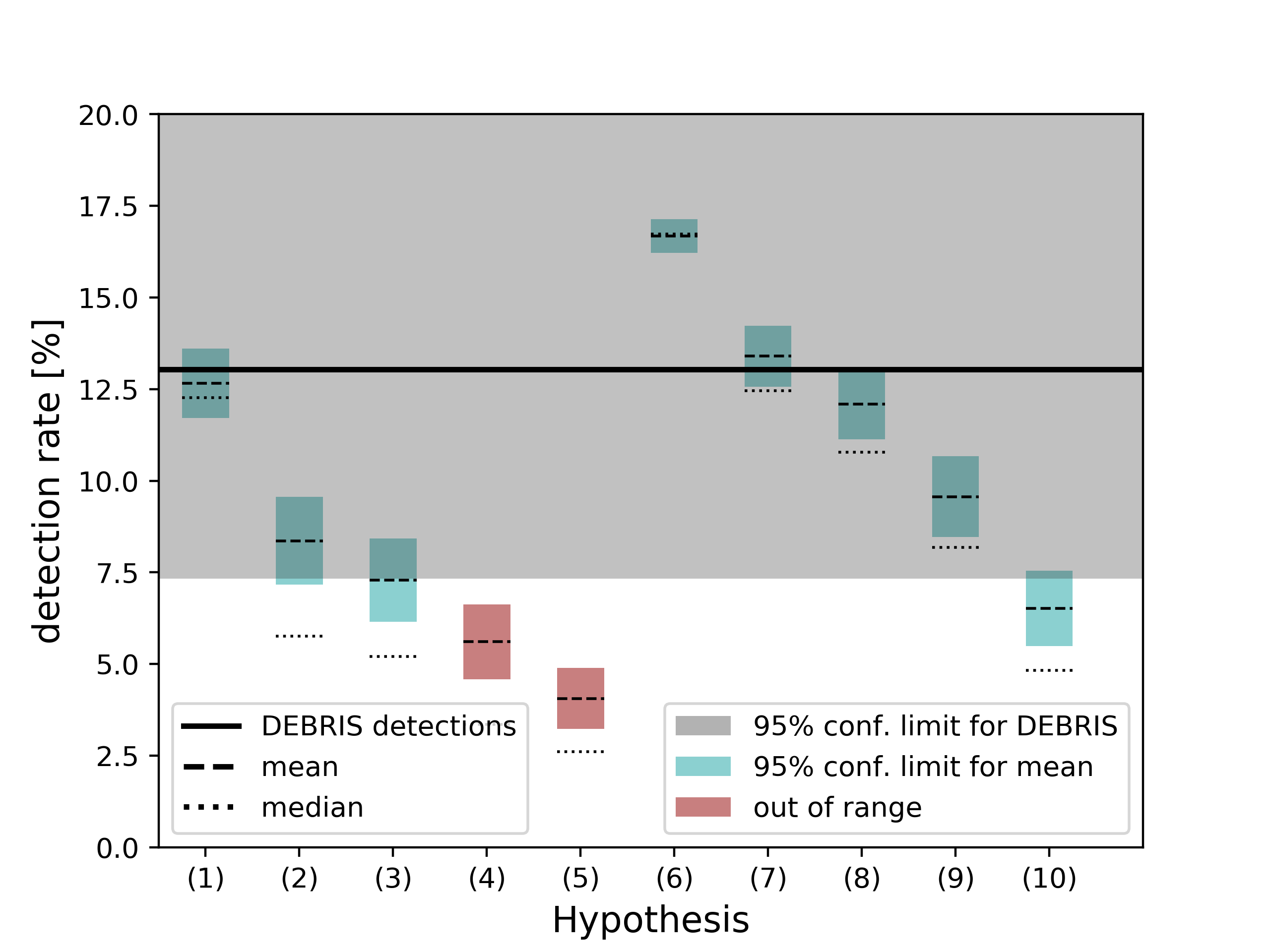}
    \end{center}
    \caption[The LOF caption]{\label{fig:AFG-to-K-statistic} Test of AFG-to K-type stars.
The black horizontal line is the DEBRIS detection rate of debris discs around K-stars. The gray-filled area is the binomial confidence interval of this detection rate at a 95\% confidence level (Jeffreys' interval\footnotemark[2]). Each hypothesis predicts individual detection rates for all 92 K-stars, $p_i$. The vertical bars show the 95\% confidence level of the mean of those $p_i$. The dashed lines on these bars show the mean and the dotted lines show the median for every hypothesis. For the detection rate of the reference stars we use the combined detection rate of A-, F- and G-stars, which is 20.8\%.}
\end{figure}

We first tested our method described in section~\ref{chap:methods} by considering the A-, F-, and G-stars and using them to predict detections of discs around K-stars in the DEBRIS sample.  
Specifically, we took 56 AFG-stars with debris discs as reference ones and generated a population of debris discs expected around K-stars from \citet{sibthorpe-et-al-2018}. We compared the predicted detection rate with the actual one. In Fig.~\ref{fig:AFG-to-K-statistic}, the predicted mean detection rate for every hypothesis with its 95\% confidence level is plotted. This has to be compared to the DEBRIS detection rate of the K-star discs, which is also shown, together with its binomial uncertainty interval (at the 95\% confidence level). A comparison suggests that hypotheses 1, 7 and 8 provide the best matches. However, all besides hypotheses 4 and 5 are statistically consistent with the detection range of K-star debris discs. This test shows that our method and all but two of our hypotheses work for our purposes. For our further analyses we therefore exclude hypotheses 4 and 5, but for completeness they are still shown in our prediction plots.

 \footnotetext[2]{\url{http://epitools.ausvet.com.au}}

\begin{figure}
     \begin{center}
            \hspace*{-5mm}\includegraphics[width=0.55\textwidth]{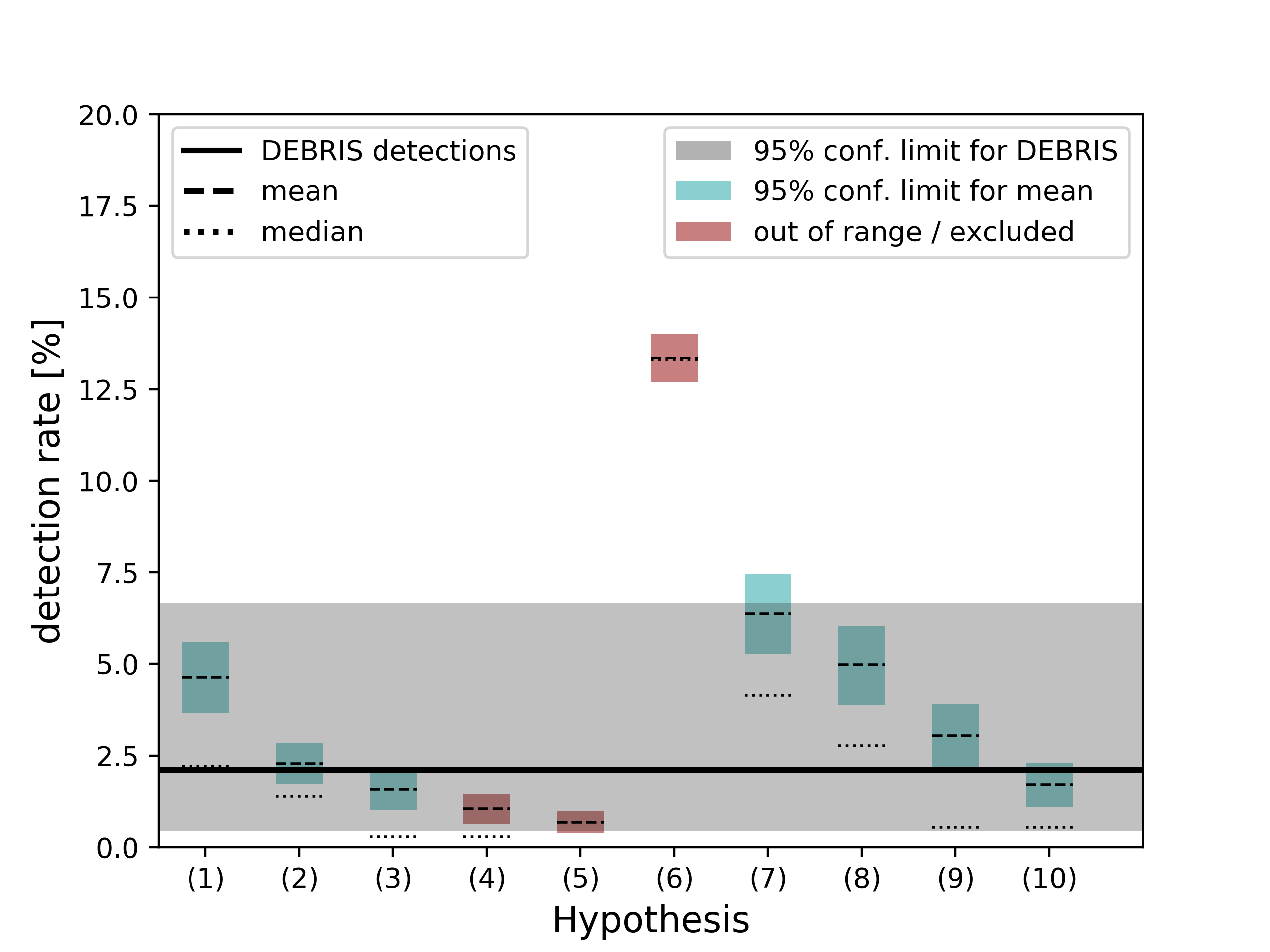}
    \end{center}
    \caption{\label{fig:DEBRIS-statistic} Detection rates of M-star debris discs. The structure of this figure is analogous to Fig.~\ref{fig:AFG-to-K-statistic}. The black horizontal line is the DEBRIS detection rate of debris discs around M-stars. Each hypothesis predicts individual detection rates for all 94 M-stars, $p_i$. Plotted is the mean of these individual rates for every hypothesis. The detection rate of the reference stars is now the combined detection rate of A-, F-, G-, and K-stars, which is 18.8\%. All hypotheses except for hypothesis~6 confirm the detection rate. 
    }
\end{figure}

\subsection{Application of hypotheses to the M-stars}

Next, we applied our method described in section~\ref{chap:methods} by considering the A-, F-, G-, K- and  M-stars in the DEBRIS sample (Table~\ref{tab:DEBRIS-sample+value-sources}). We took all 68 AFGK-stars with debris discs as reference ones and generated a population of debris discs expected around M-stars (Lestrade et al., in prep.). The predicted mean detection rate for every hypothesis along with its 95\% confidence level is plotted in Fig.~\ref{fig:DEBRIS-statistic}. As before, we overplot the actual DEBRIS detection rate with its 95\% confidence interval. A comparison demonstrates that all but hypothesis~6 are consistent with the M-star DEBRIS detection rate. We therefore also exclude hypothesis~6 from our further analyses.
We get a detection rate between 1.6\%\textpm 0.6\% and 6.4\%\textpm1.1\% for the M-stars in the DEBRIS sample. We see that the mean and the median for most hypotheses differ widely. This shows the large asymmetric spread between individual values.

This demonstrates that it is indeed possible that M-stars harbour populations of debris discs similar to those around earlier-type stars, and that many of these discs may have just eluded detection due to occupying a different region in the $T_{\fracd}$ vs. $f_{\fracd}$ plot compared to discs around earlier type stars. Unfortunately, with the \emph{Herschel} data we can only rule out three of the hypotheses. Surveys with other facilities will be necessary to distinguish between the remaining ones.


\subsection{Prediction for IRAM surveys}
\label{subsec:IRAMgalaxies}

\begin{figure}
     \begin{center}
       \includegraphics[width=0.5\textwidth]{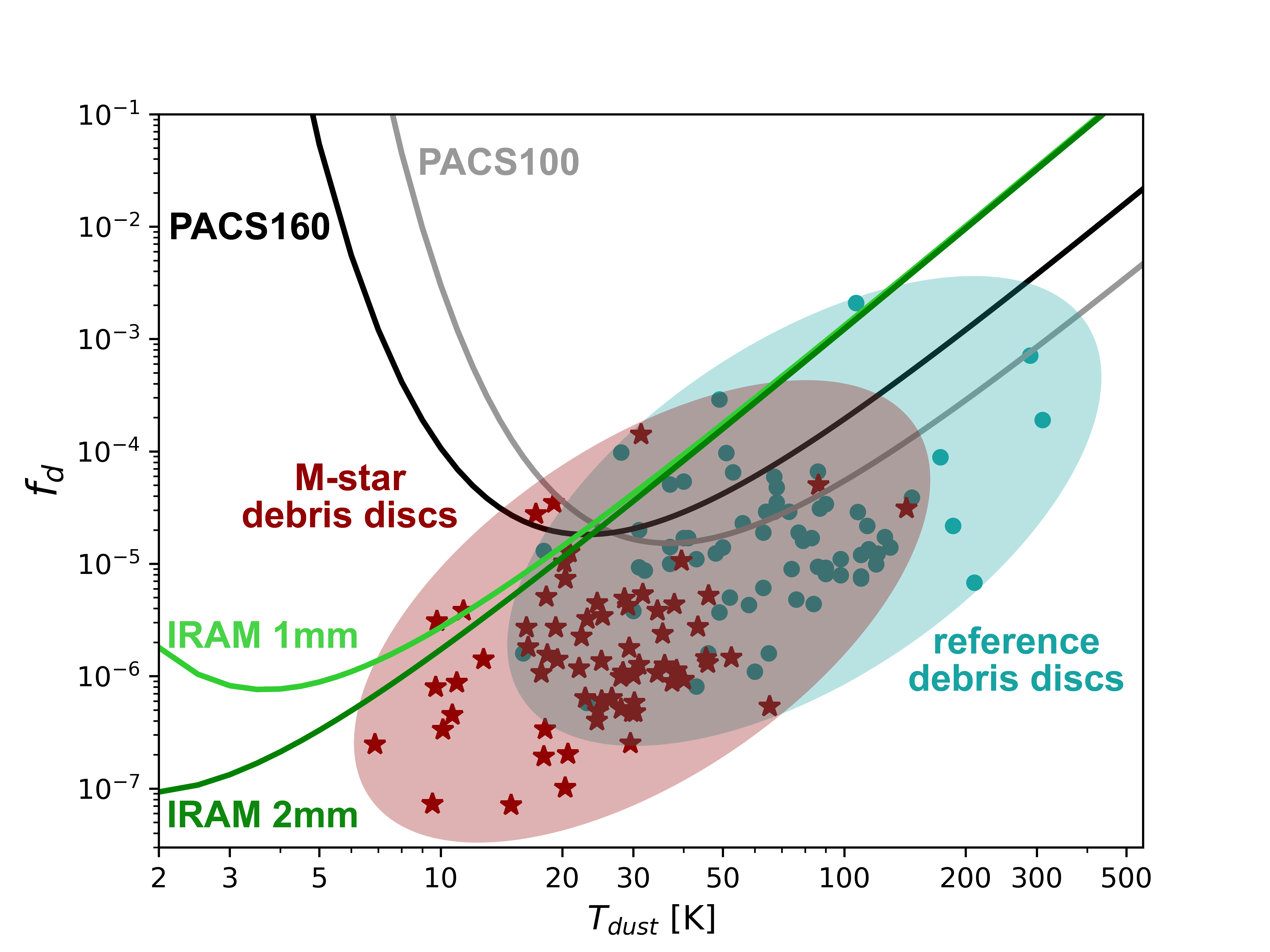}
    \end{center}
    \caption{Same as Fig.~\ref{fig:schematic}, but with the 15~min IRAM detection limits overplotted. The light green curve represents the IRAM 1~mm band and the dark green curve shows the IRAM 2~mm band.
}
    \label{fig:IRAM-shape}
\end{figure}

Next, we predict the number of detections for the M-stars from the DEBRIS survey that would be expected if this sample were observed with the new NIKA-2 instrument on IRAM \citep{adam-et-al-2018, catalano-et-al-2016,catalano-et-al-2018}. To this end, we do the same calculation as before, but change the detection limit curves to those of the 1~mm and 2~mm bands of NIKA-2. A comparison of the detection limit curves is shown in Fig.~\ref{fig:IRAM-shape}. We assumed different observing times of 15~min, 30~min and 60~min and therefore different sensitivities, see Table~\ref{tab:IRAM_table}.

\begin{table}
	\centering
	\caption{IRAM sensitivities in different bands for different observation times.}
	\label{tab:IRAM_table}
	\begin{tabular}{lcccl}
		\hline
		 & & 15~min & 30~min & 60~min \\
		\hline
		  1~mm & 3$\sigma$ 	& 3.3 mJy & 2.3 mJy & 1.65 mJy \\
		  2~mm & 3$\sigma$ 	& 0.8 mJy & 0.6 mJy & 0.4 mJy \\
		\hline
	\end{tabular}
			\\[3mm]
\textbf{Note:}
The sensitivities are calculated from those in Kramer \& Sanchez Portal 2019\footnotemark[3]
\end{table}
\footnotetext[3]{\url{http://www.iram.fr/GENERAL/calls/s19/30mCapabilities.pdf}}

\begin{figure*}
     \centering
     \begin{subfigure}[t]{0.3435\textwidth}
         \centering
         \includegraphics[width=\textwidth]{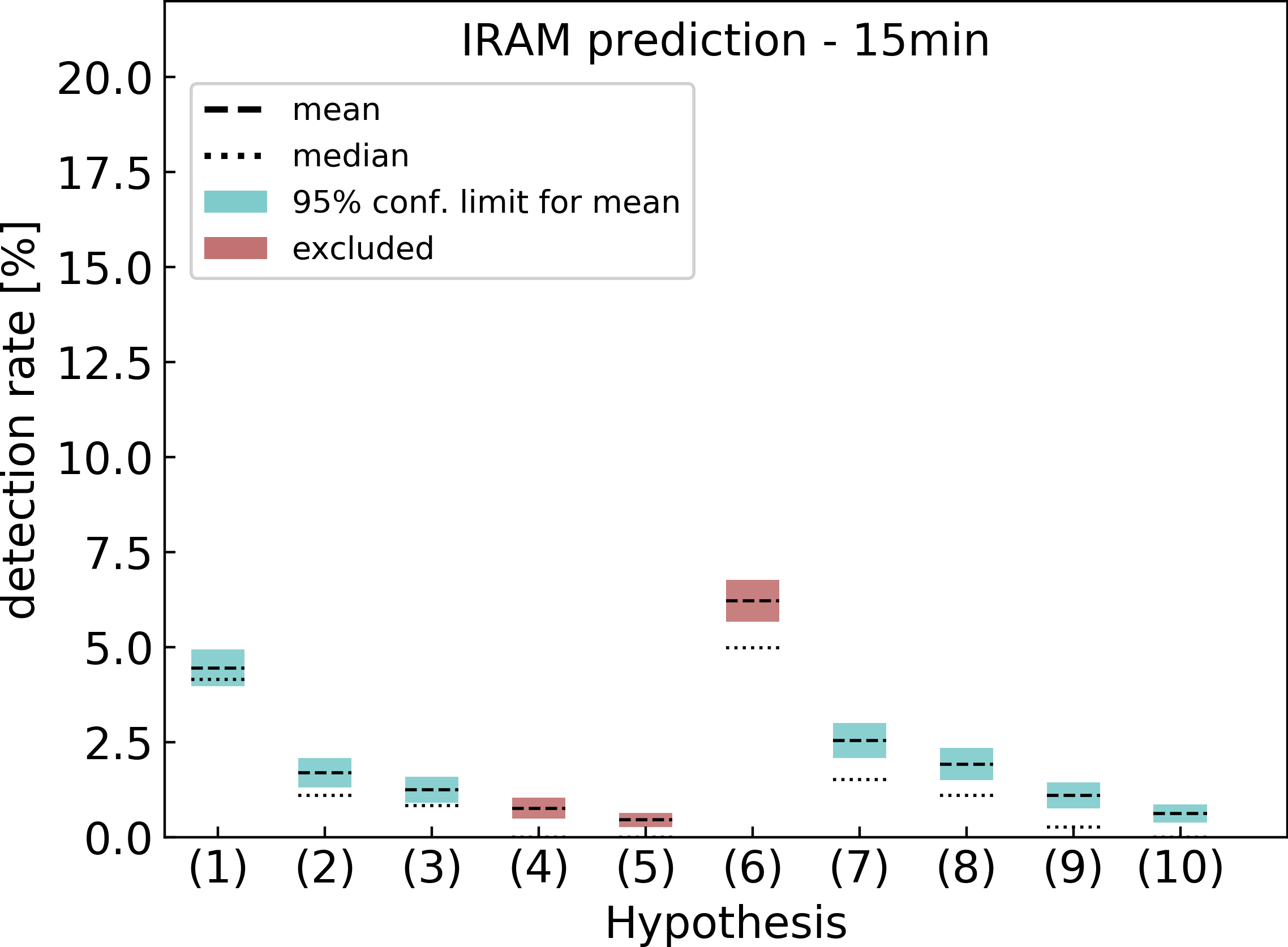}
     \end{subfigure}
     \begin{subfigure}[t]{0.3\textwidth}
         \centering
         \includegraphics[width=\textwidth]{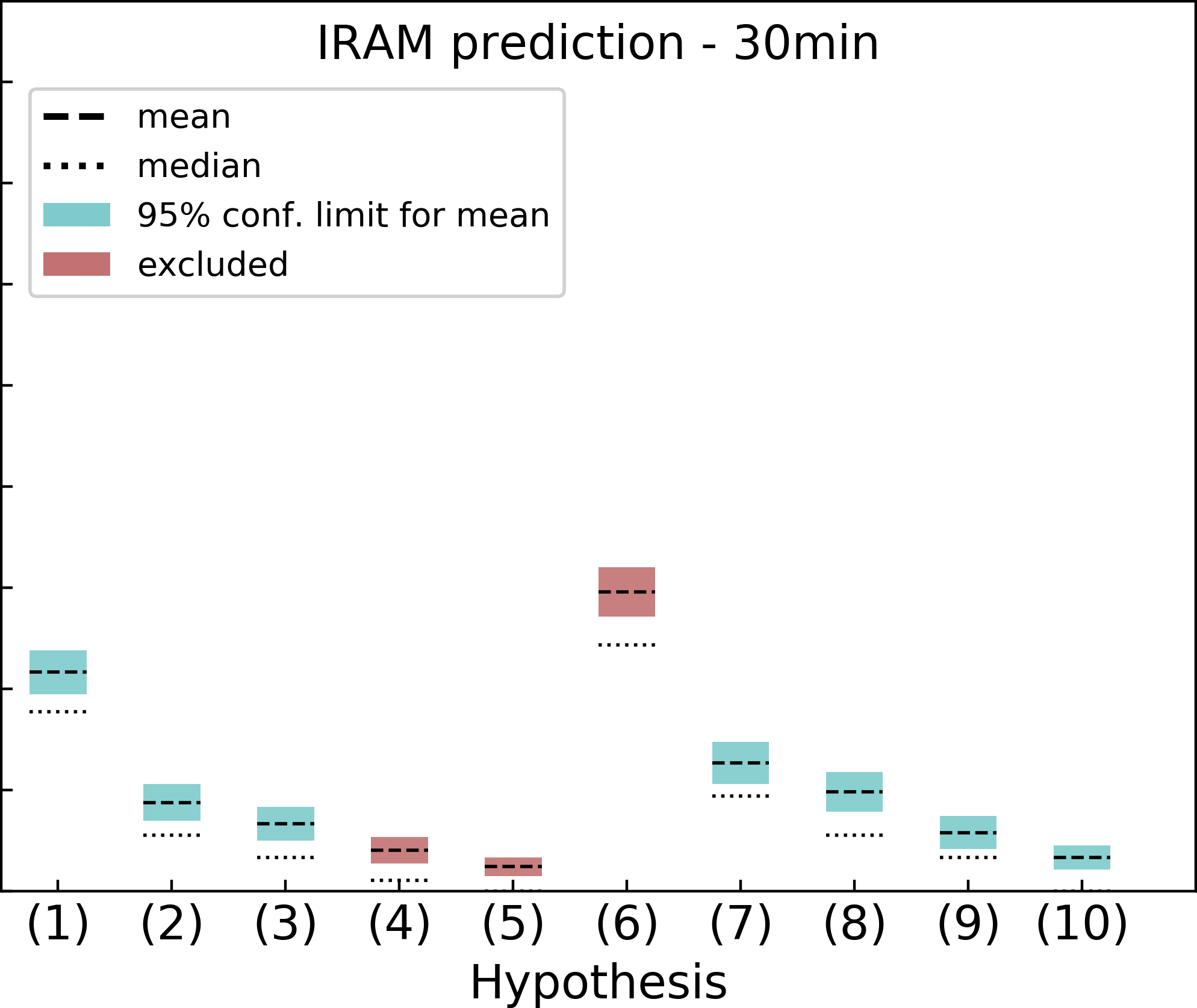}
     \end{subfigure}
     \begin{subfigure}[t]{0.3\textwidth}
         \centering
         \includegraphics[width=\textwidth]{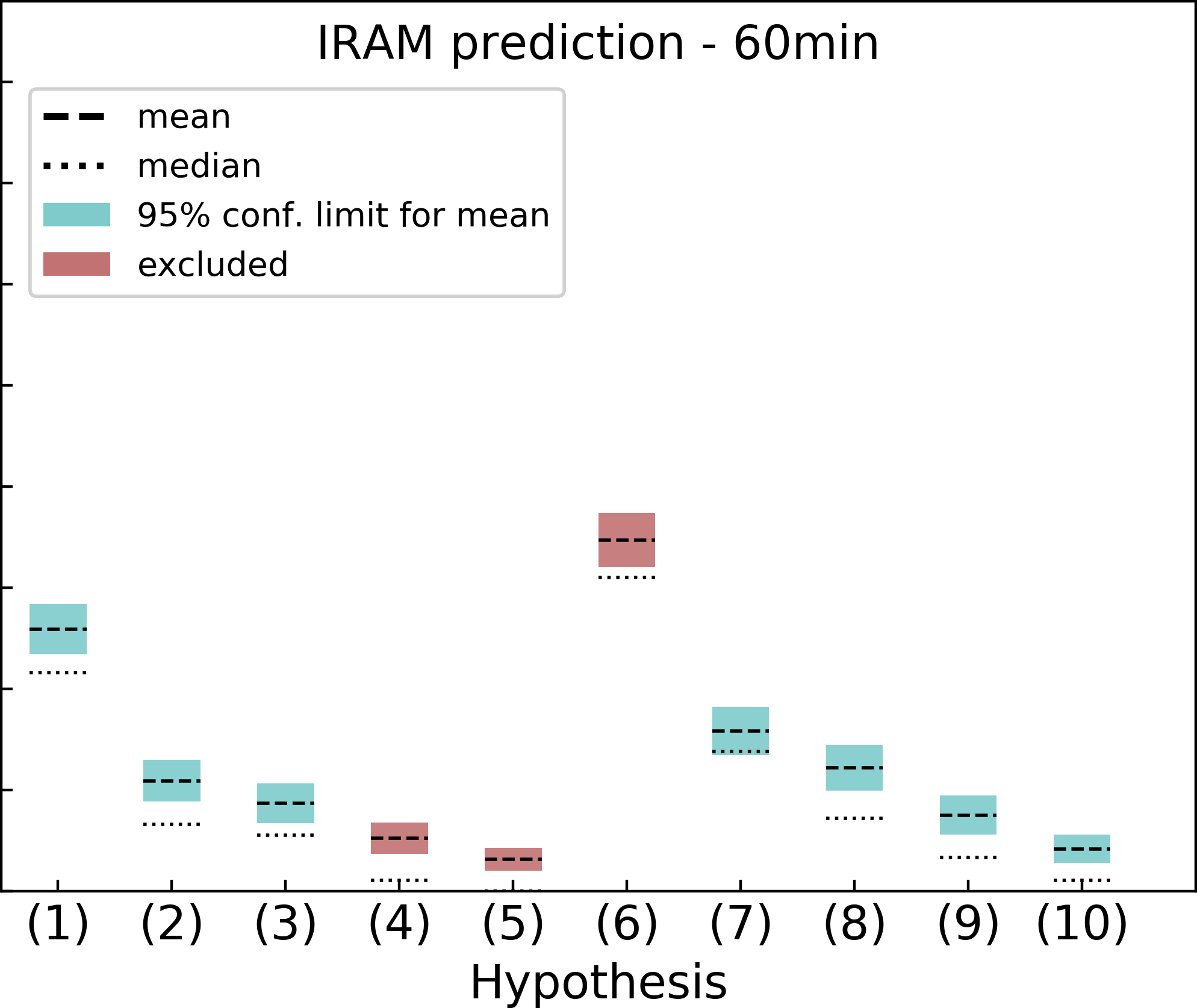}
     \end{subfigure}
        \caption{Prediction for an observation with IRAM with three different observing times.}
        \label{fig:IRAM-prediction}
\end{figure*}

In Fig.~\ref{fig:IRAM-prediction} we plot the detection rates for the different observing times and hypotheses. The lowest and highest rates, including 95\% confidence limits, are listed in Table~\ref{tab:IRAM_detection_prob}. As expected, the minimum and maximum detection rates get higher for longer observing times.

If we compare the 15~min observation of the DEBRIS survey (Figure~\ref{fig:DEBRIS-statistic}) with the 15~min observation of IRAM (Figure~\ref{fig:IRAM-prediction} left), we see a slightly lower detection rate for the observation with IRAM.
The conclusion is that the NIKA-2 IRAM instrument would not provide a better detection rate than \emph{Herschel}/PACS did. However, IRAM could detect colder and fainter M-star discs than PACS, which would result in partly different disc detections (see Fig.~\ref{fig:IRAM-shape}).

A potential issue with these proposed IRAM observations is the possibility of contamination by background galaxies. To calculate the number of galaxies above a given flux density expected within a certain area we used the Schecter function as shown in \citet{booth-et-al-2017}:

\begin{equation}
    n(F_{\nu}) \fracd \, \fraclog \, F_{\nu} = A \phi_{*} 
    \left( \frac{F_{\nu}}{S_{*}} \right)^{\alpha+1} 
    \exp \left( \frac{-F_{\nu}}{S_{*}} \right)
    \natlog \, 10 \,
    \fracd \, \fraclog \, F_{\nu} .
	\label{eq:schechter function}
\end{equation}

Parameter $A$ is the area of sky in deg$^{2}$ and $\phi_{*}$, $S_{*}$ and $\alpha$ are parameters that depend on the wavelength of the observations. These parameters are determined through surveys for galaxies. Here we will use the values provided by \citet{carniani-et-al-2015} for observations at 1.3~mm. We therefore need to adjust our flux densities to the appropriate wavelength. A galaxy SED slope is $F_{\nu} \propto \lambda^{-2}$ modified by an additional factor $\left(\lambda/\lambda_{0}\right)^{-\beta}$, with $\beta=-1.6$ \citep{casey-2012}. Accordingly, we assumed a slope of $\lambda^{-3.6\pm 0.38}$ to translate our 1 and 2~mm flux densities  to flux densities at 1.3~mm.

Using this we then considered the case of a 60~min IRAM observation assuming a >3$\sigma$ detection. We determine that there is a 6.6\% and 1.8\% chance of finding a galaxy within 1 beam of a star for IRAM 1~mm and 2~mm, respectively. These probabilities are similar to the probabilities of detecting a debris disc around an M-star thus meaning that an IRAM survey for debris discs around M-stars is unfortunately likely to detect as many galaxies coincident with the stars as it does detect debris discs around them.

\begin{table}
	\centering
	\caption{IRAM minimum and maximum detection probabilities}
	\label{tab:IRAM_detection_prob}
	\begin{tabular}{lcl}
		\hline
		 & min [\%] & max [\%] \\
		\hline
		    15~min	& 0.6 \textpm 0.2 & 4.5 \textpm 0.5 \\
		    30~min	& 0.8 \textpm 0.3 & 5.4 \textpm 0.5  \\
		    60~min	& 1.0 \textpm 0.3 & 6.5 \textpm 0.6 \\
		 \hline
	\end{tabular}
			\\
\end{table}


\subsection{Prediction for future ALMA surveys}

\begin{figure}
     \begin{center}
       \includegraphics[width=0.5\textwidth]{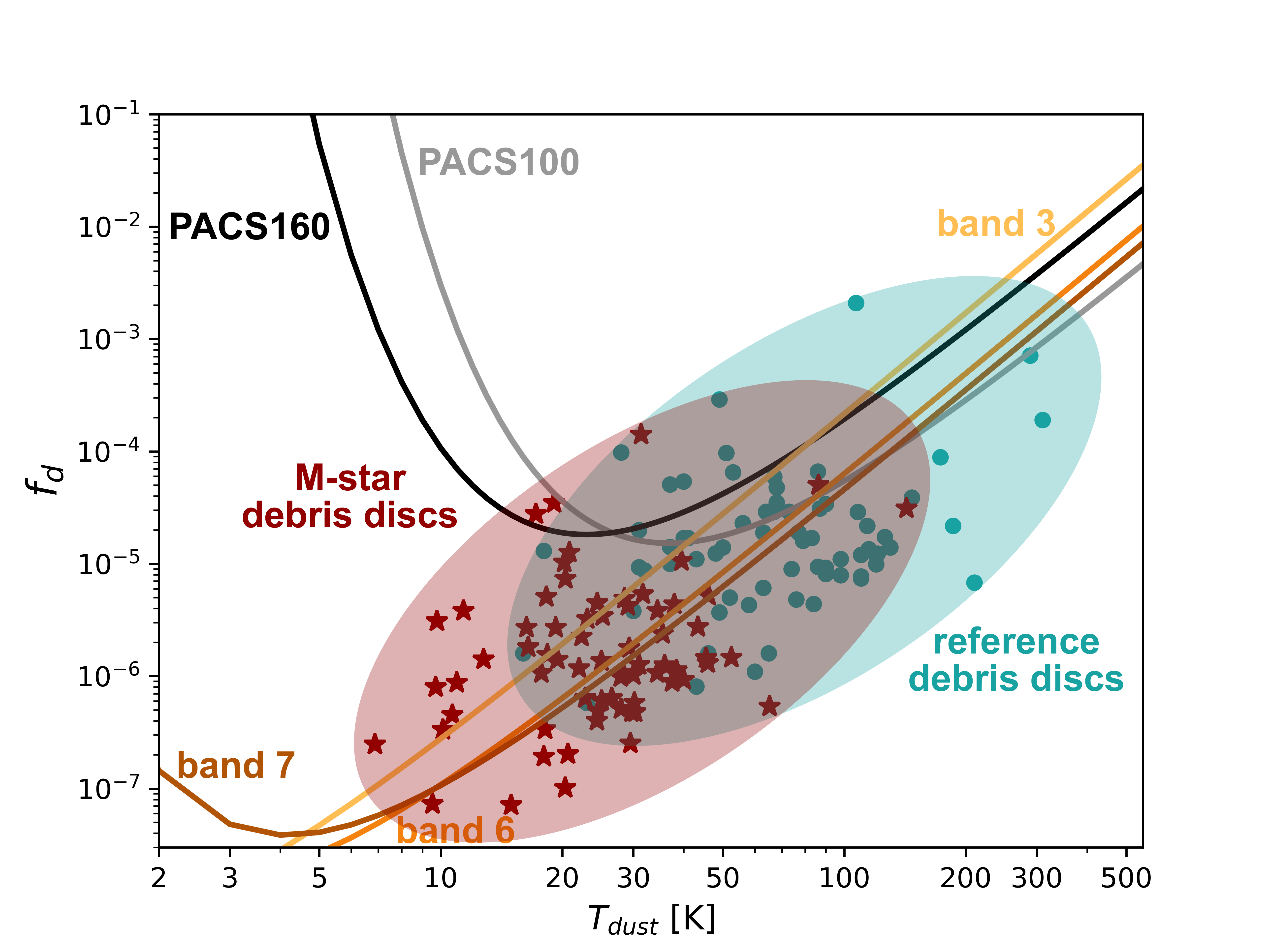}
    \end{center}
    \caption{Same as Fig.~\ref{fig:schematic}, but with the 15~min ALMA detection limits overplotted.
The light brown curve represents ALMA band 3, brown curve ALMA band 6 and the dark brown curve shows ALMA band 7. 
}
    \label{fig:ALMA-shape}
\end{figure}
We now assume that the same DEBRIS sample of M-stars (Lestrade et al., in prep.) is observed with ALMA in a dedicated program. We consider observations with the 12m array in its most compact configuration in bands 3 (111 GHz), 6 (226 GHz) and 7 (340 GHz). A comparison of the detection limits from DEBRIS and ALMA is shown in Figure~\ref{fig:ALMA-shape}. The sensitivities are listed in Table~\ref{tab:ALMA_table}.

\begin{table}
	\centering
	\caption{ALMA sensitivities in different bands for different observation times.}
	\label{tab:ALMA_table}
	\begin{tabular}{lcccl}
		\hline
		 & & 15~min & 30~min & 60~min \\
		\hline
		  Band 3 & 3$\sigma$ 	& 78.3 $\mu$Jy & 55.4 $\mu$Jy & 39.2 $\mu$Jy \\
		  Band 6 & 3$\sigma$ 	& 92.4 $\mu$Jy & 65.4 $\mu$Jy & 46.2 $\mu$Jy \\
		  Band 7 & 3$\sigma$ 	& 147.4 $\mu$Jy & 104.2 $\mu$Jy & 73.7 $\mu$Jy \\
		\hline
	\end{tabular}
	\\[3mm]
\textbf{Note:}
The sensitivities were calculated with the ALMA sensitivity calculator\footnotemark[4]
\end{table}
\footnotetext[4]{\url{https://almascience.eso.org/proposing/sensitivity-calculator}}

The lowest and highest detection rates for bands 3, 6 and 7 for observing times of 15~min, 30~min and 60~min, including 95\% confidence limits, are given in the left part of Table~\ref{tab:ALMA_detection_prob}.
The detection rates in bands 6 and 7 do not differ much. 

\begin{figure*}
     \centering
     \begin{subfigure}[b]{0.34345\textwidth}
         \centering
         \includegraphics[width=\textwidth]{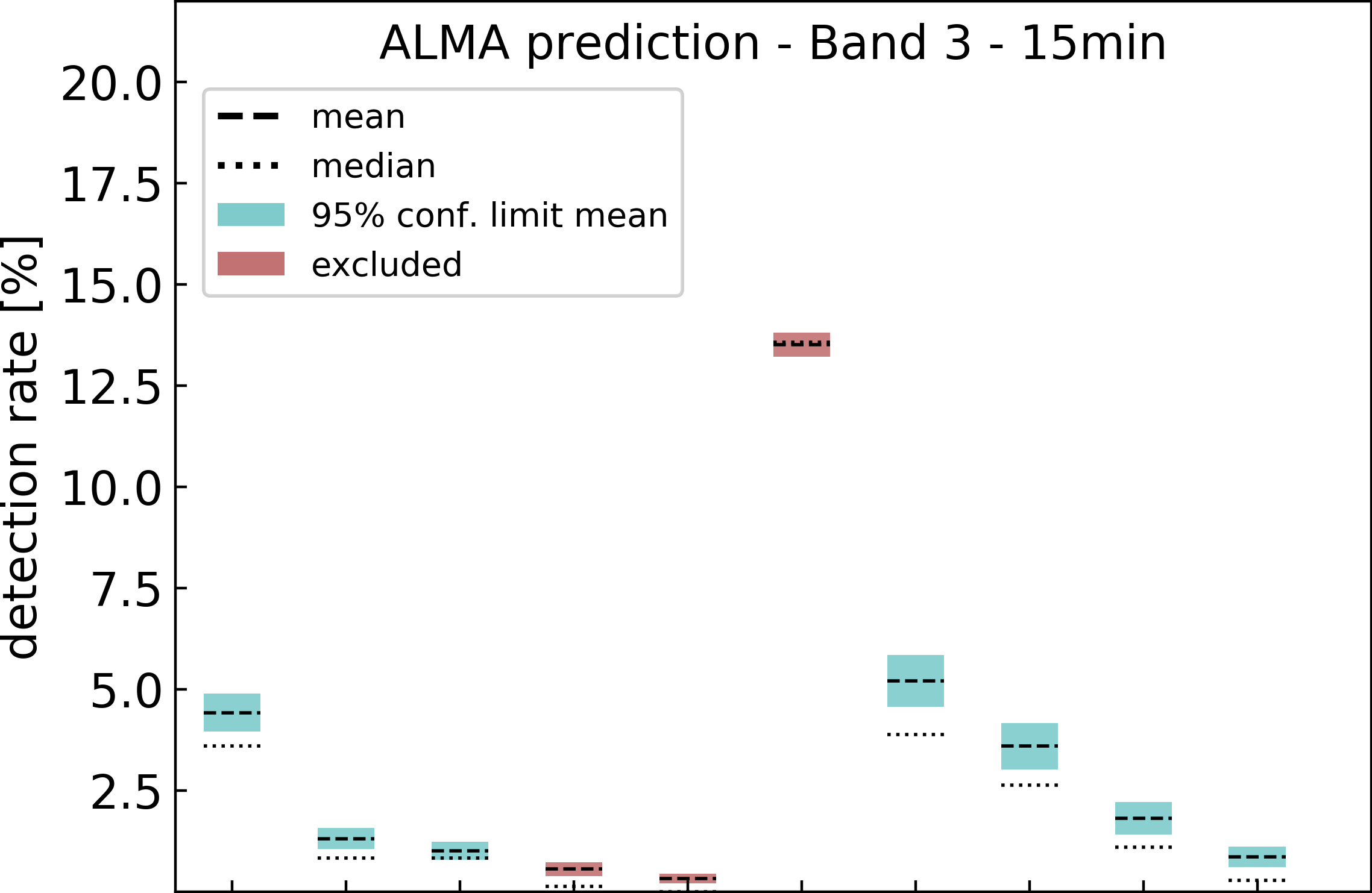}
		 \includegraphics[width=\textwidth]{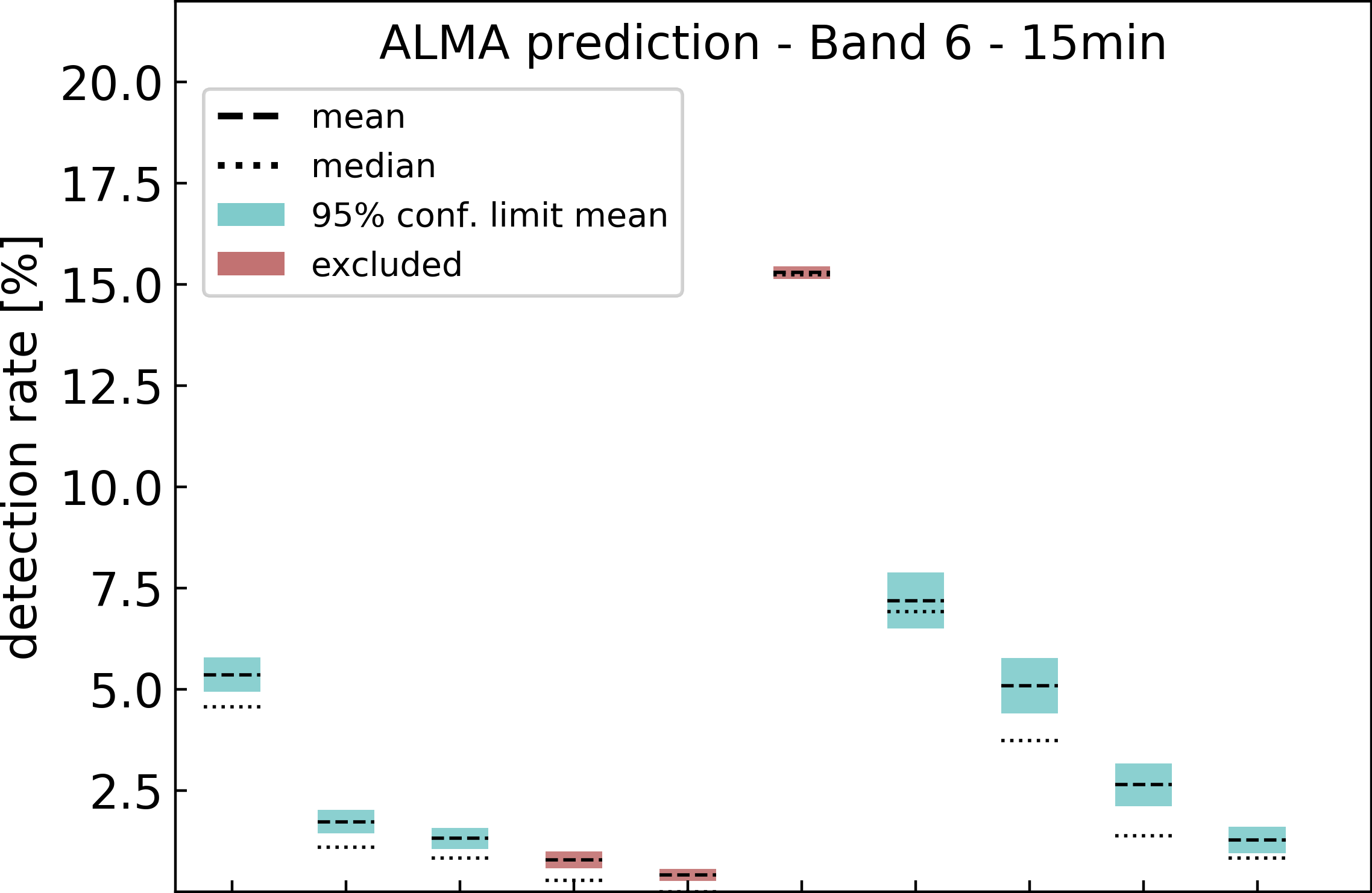}
         \includegraphics[width=\textwidth]{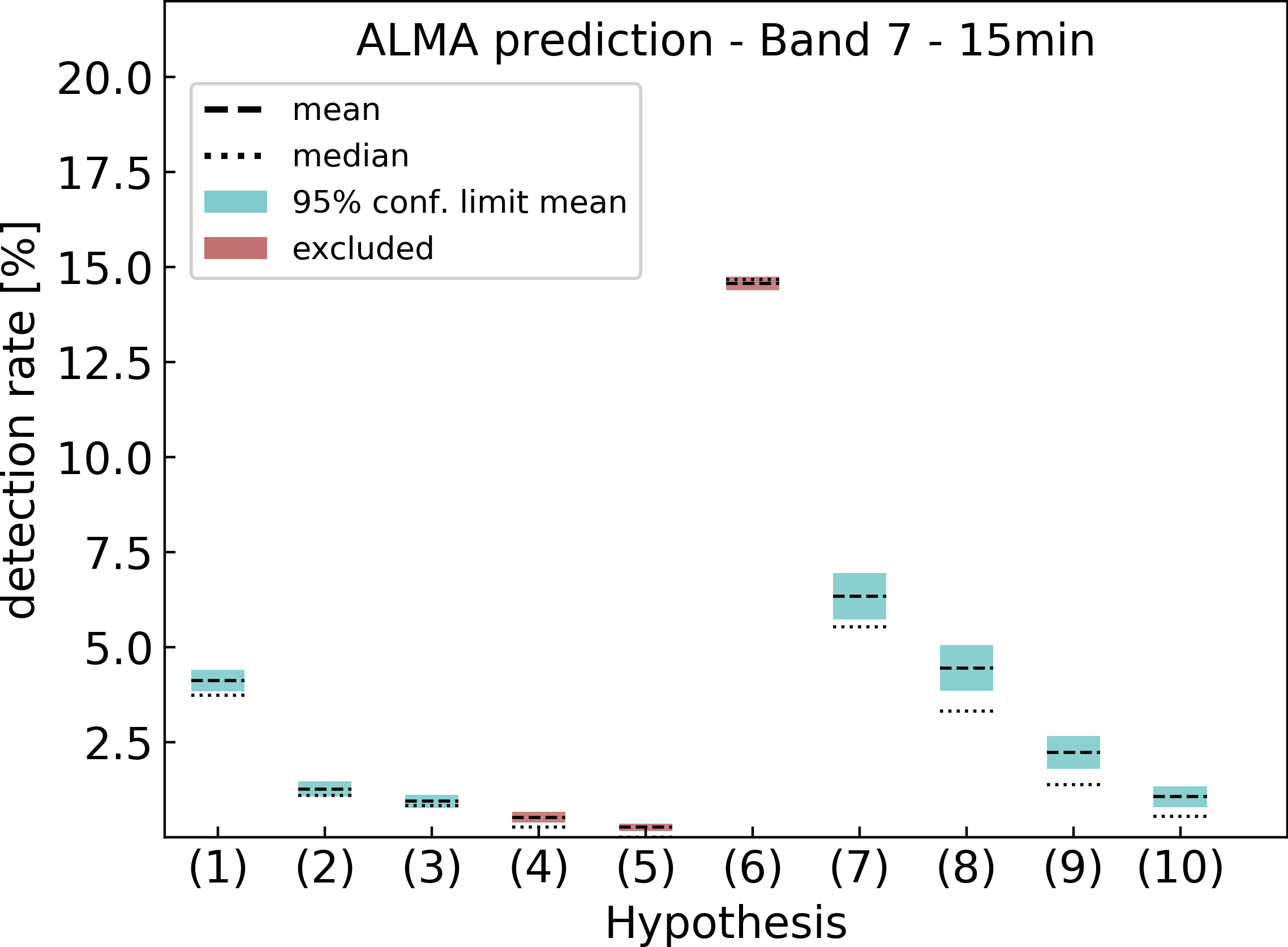}
     \end{subfigure}%
     \begin{subfigure}[b]{0.3\textwidth}
         \centering
         \includegraphics[width=\textwidth]{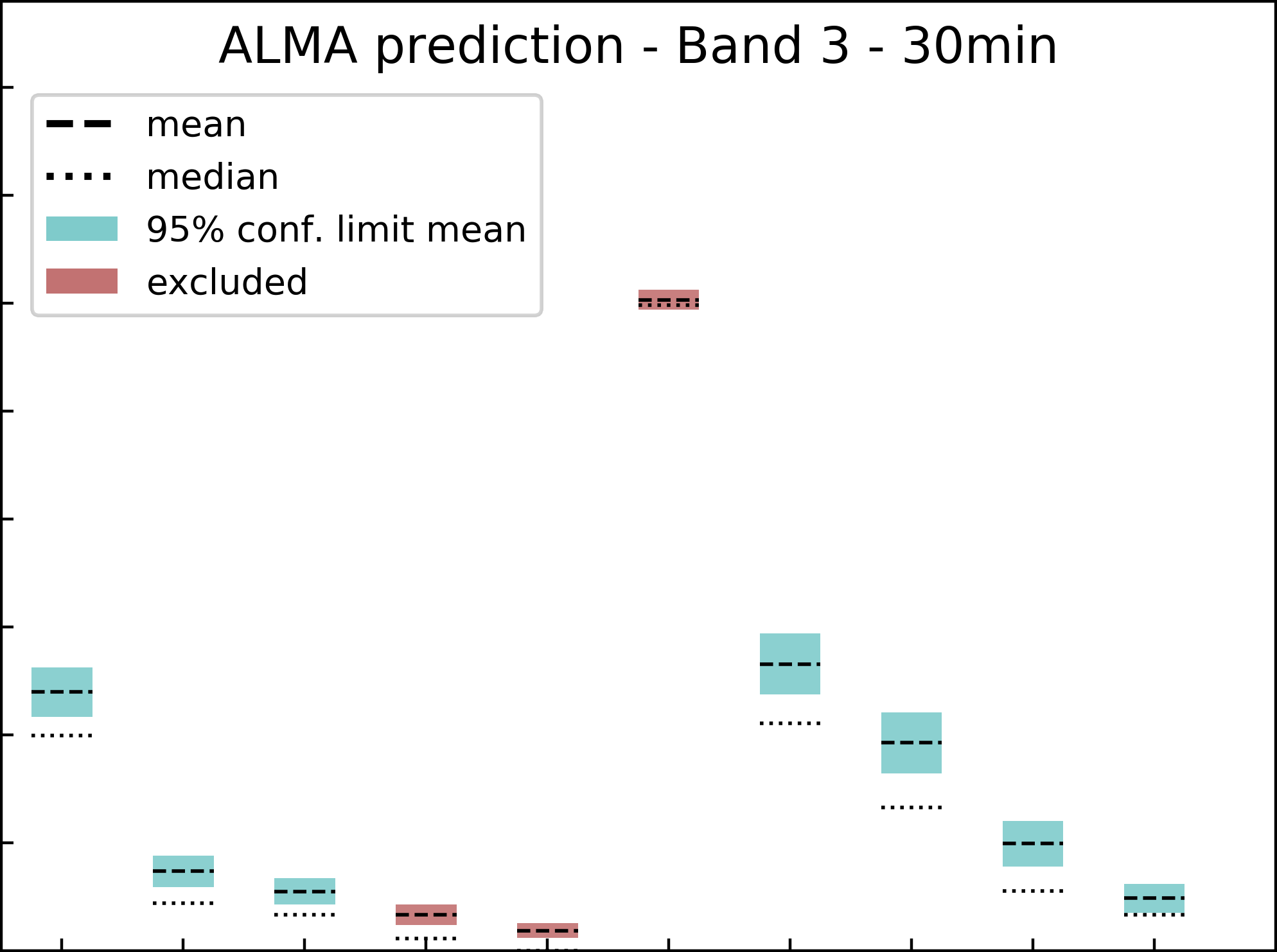}
         \includegraphics[width=\textwidth]{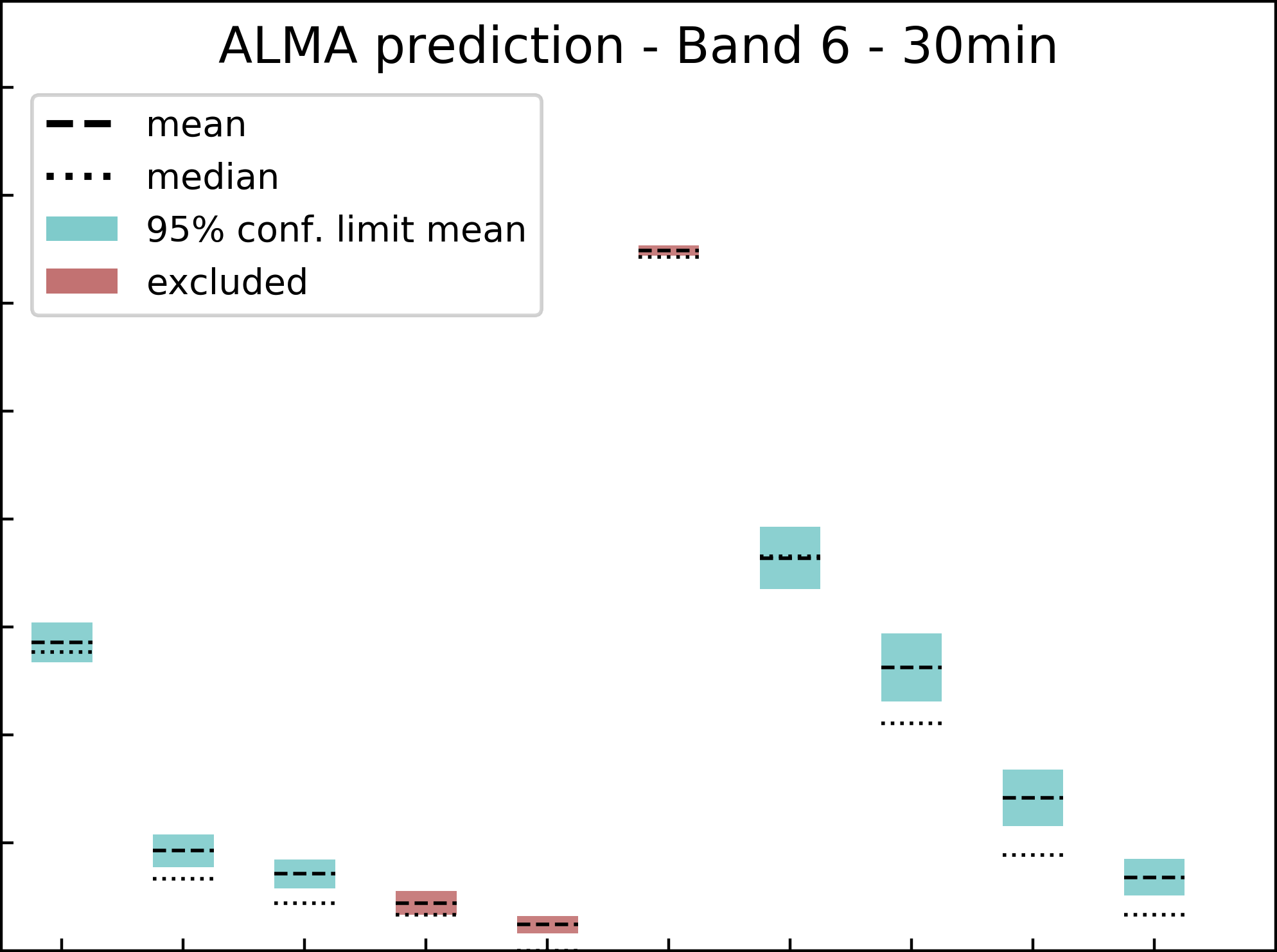}
         \includegraphics[width=\textwidth]{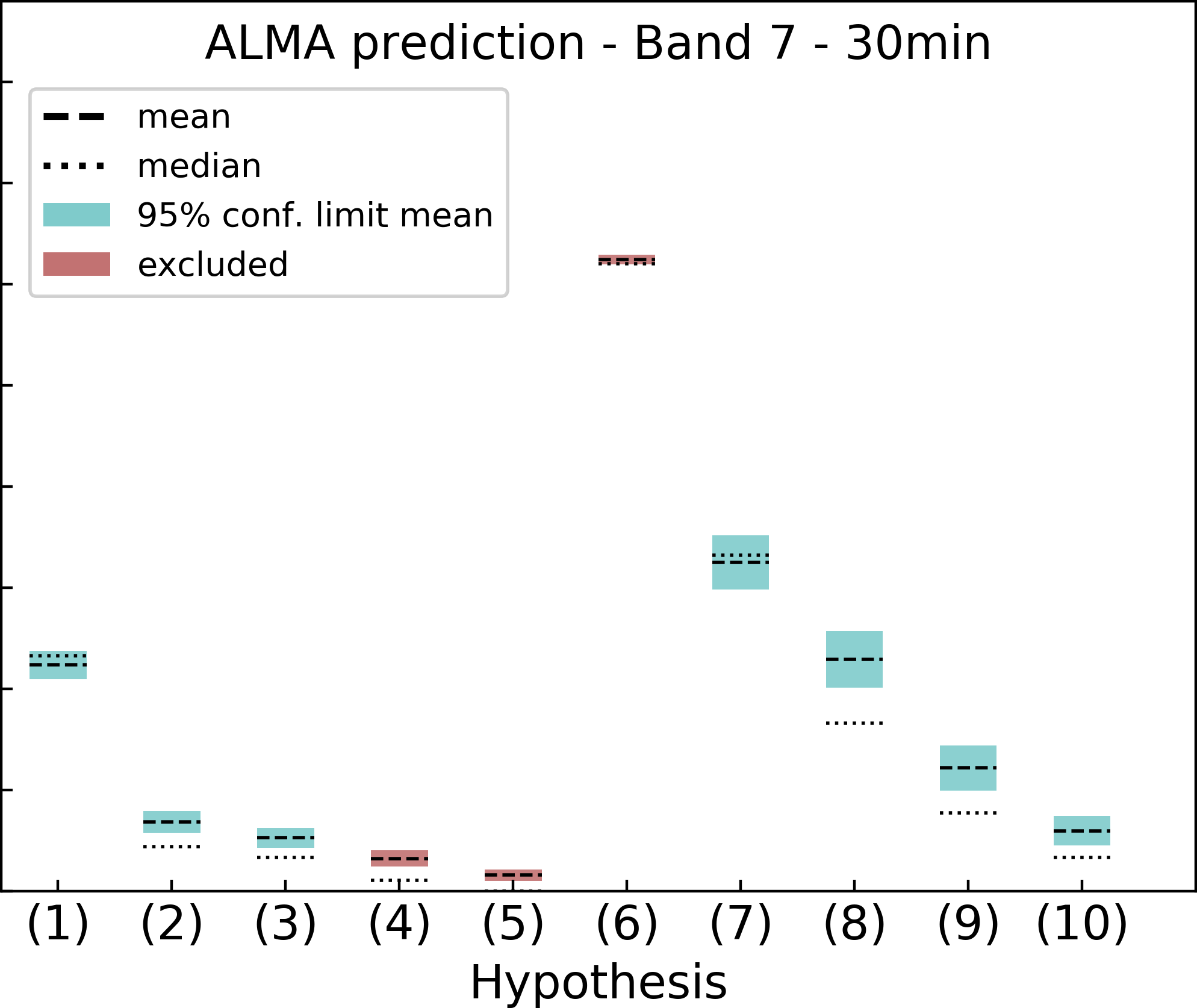}
     \end{subfigure}%
     \begin{subfigure}[b]{0.3\textwidth}
         \centering
         \includegraphics[width=\textwidth]{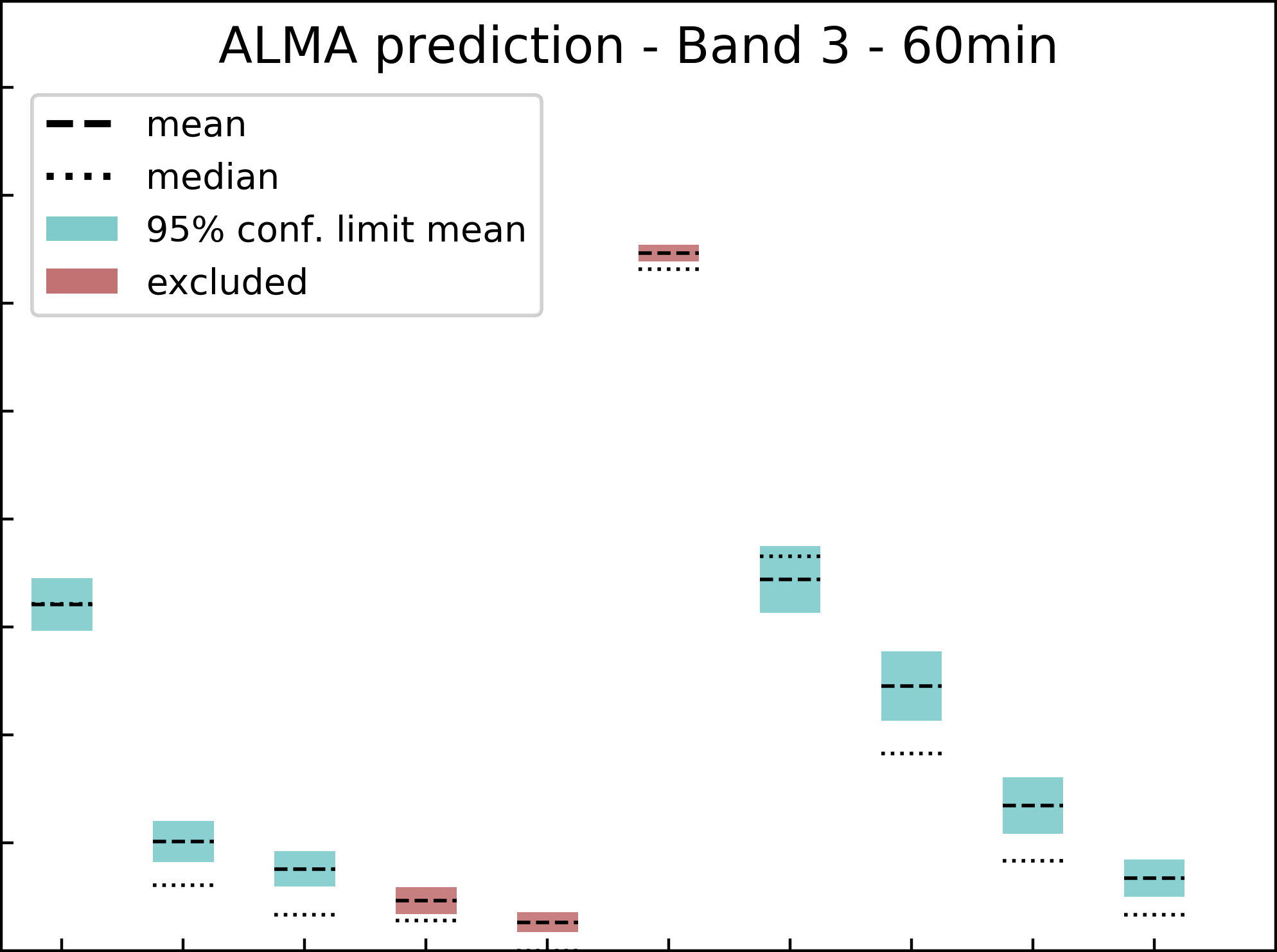}
         \includegraphics[width=\textwidth]{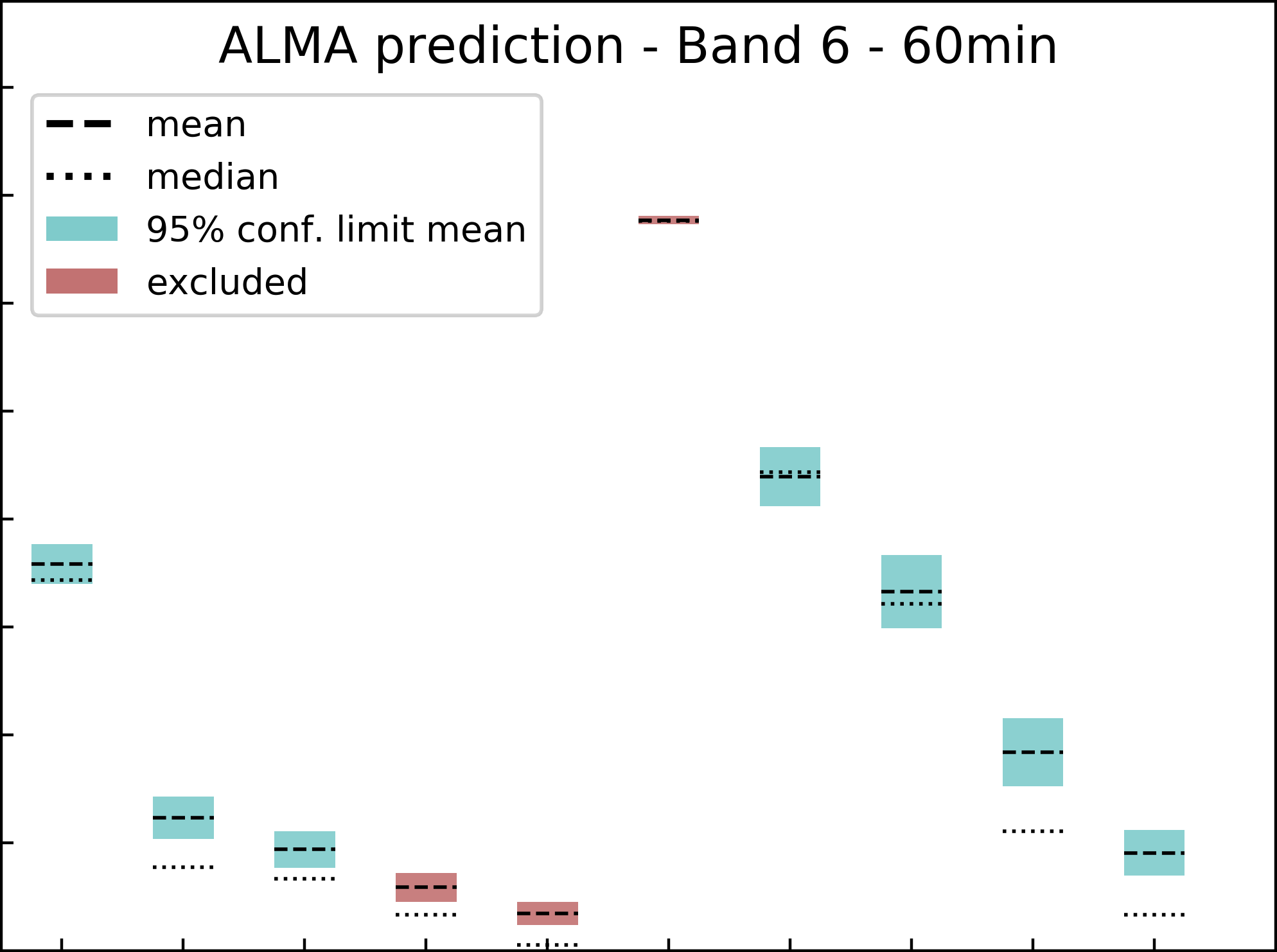}
         \includegraphics[width=\textwidth]{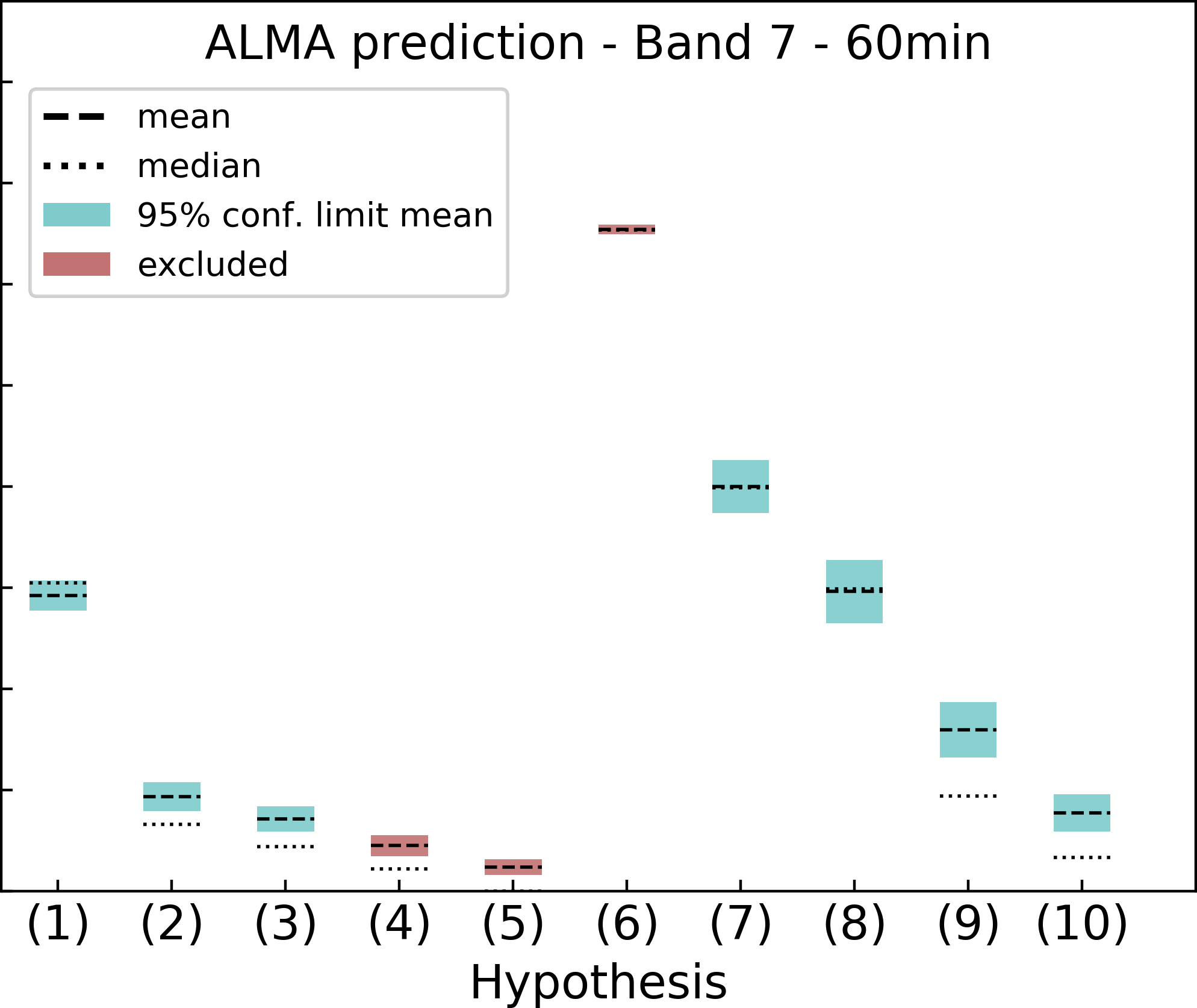}
     \end{subfigure}%
        \caption{Prediction for an observation with ALMA Band 3, 6 and 7 with three different observing times each.}
        \label{fig:ALMA-band-3,6,7-prediction}
\end{figure*}

\begin{table*}
	\centering
	\caption{ALMA minimum and maximum detection probabilities firstly considering only the total flux of the disc and secondly taking into account the reduction of the flux per beam due to many of the discs being resolved.}
	\label{tab:ALMA_detection_prob}
	\begin{tabular}{lccc|cc|cl}
	    \hline
		 & & \multicolumn{2}{c}{using total disc flux}
	     & \multicolumn{2}{c}{correcting for the resolution}\\
		 & & min [\%] & max [\%] & min [\%] & max [\%] \\
		\hline
		  Band 3 & 15~min	& 2.1 \textpm 0.6 & 10.5 \textpm 0.7 
		  			& 0.9 \textpm 0.2  & 5.2 \textpm 0.6 \\ 
		         & 30~min	& 2.6 \textpm 0.6 & 12.0 \textpm 0.7 
		         	& 1.2 \textpm 0.3  & 6.6 \textpm 0.7 \\ 
		         & 60~min	& 3.1 \textpm 0.7 & 13.4 \textpm 0.7 
		         	& 1.7 \textpm 0.4  & 8.6 \textpm 0.8 \\ 
		 \hline
		  Band 6 & 15~min	& 3.9 \textpm 0.8 & 15.1 \textpm 0.6 
		  			& 1.3 \textpm 0.3  & 7.2 \textpm 0.7 \\ 
		         & 30~min	& 4.5 \textpm 0.9 & 16.1 \textpm 0.5 
		         	& 1.7 \textpm 0.4  & 9.1 \textpm 0.7 \\ 
		         & 60~min	& 5.3 \textpm 1.0 & 17.1 \textpm 0.4 
		         	& 2.3 \textpm 0.5  & 11.0 \textpm 0.7 \\ 
		\hline
		  Band 7 & 15~min	& 4.3 \textpm 0.9 & 15.8 \textpm 0.5 
		  			& 1.0 \textpm 0.2  & 6.3 \textpm 0.6 \\ 
		         & 30~min	& 5.1 \textpm 1.0 & 16.8 \textpm 0.4 
		         	& 1.3 \textpm 0.2  & 8.1 \textpm 0.7 \\ 
		         & 60~min	& 5.9 \textpm 1.1 & 17.6 \textpm 0.3 
		         	& 1.8 \textpm 0.3  & 10.0 \textpm 0.7 \\ 
		\hline
	\end{tabular}
			\\
\end{table*}

\subsubsection{Resolved discs}
In the previous section we have made the assumption that the discs are unresolved. However, ALMA not only has a superior sensitivity to previous millimetre observatories, but also a superior resolution. This makes it an excellent observatory for detailed studies of the structure of debris discs, but may prove to be problematic for our attempts to detect discs around the nearest M-stars. Their proximity means the discs have a large angular size, which will result in a decreased flux density per beam. In order to understand how strongly this affects our results we have checked how many of our calculated debris discs have a diameter larger than the resolution for the different bands, see Table~\ref{tab:resolved_table}. Our calculation shows that the number of discs that would be resolved increases with increasing frequency. For the cases where $R\propto L_{*}^{0.19}$ (hypotheses~6 to~10) the number of resolved discs increases from band 3 with 20.4\% to 54.4\% for band 7. For $R=\mathrm{const}$ (hypotheses~1 to~5) the numbers get even higher, with 55.5\% for band 3 to 90.6\% for band 7.

To get an estimate of how the detection rates may be affected, we calculated for every band the mean number of beams that are necessary to cover the complete disc by dividing the disc circumference by the beam size. The values were determined by calculating the number of beams necessary for every single disc and dividing the fractional luminosity by the number of beams. For $R\propto L_{*}^{0.19}$ we get means of 2.4 beams for band 3, 5.0 beams for band 6 and 7.5 beams for band 7. The number increases for hypotheses with $R=\mathrm{const}$ to 6.7 beams for band 3, 13.7 beams for band 6 and 20.6 beams for band 7. These large sizes, especially for the hypotheses with $R=\mathrm{const}$, result in a reduction to the detection rates calculated in the previous section.

Since ALMA is an interferometer we also have to consider the Maximum Recoverable Scale (MRS)\footnotemark[5]. The MRS is the largest scale size that can be detected by the smallest baseline of an interferometer. This scale has to be compared with the Largest Angular Scale (LAS) of the disc. If the LAS of the disc is bigger than the MRS, then some of the emission will not be detectable. Calculating the exact effect would require time consuming simulations for every disc. We, therefore, decided to use a conservative approach and assumed that any disc with a diameter larger than the MRS is not observable and set their fractional luminosity to zero.
\footnotetext[5]{\url{https://almascience.eso.org/proposing/proposers-guide}}

In this way we got a ``corrected'' fractional luminosity for every disc, which decreased our detection rates. The minimum and maximum detection rates after accounting for this correction are compared to those simply using the total flux in Table~\ref{tab:ALMA_detection_prob}. The minimum detection rates are now all $\leq 2.3\%$ and the maximum detection rates are roughly half as high as before. Due to the high resolution of band 7, the detection rates of this band become even lower than those of band 6.
Including the MRS only had a minor effect on the detection rates since discs that are larger than the MRS will already be difficult to detect due to their low flux density per beam. For band 3 the detection rates did not change, for band 6 they decreased by 0.1--0.2\% and for band 7 they decreased by 0.2--0.5\%.

In Fig.~\ref{fig:ALMA-band-3,6,7-prediction} we plot the detection rates for the resolved discs for bands 3, 6 and 7 for observing times of 15~min, 30~min and 60~min. In contrast to Herschel and IRAM, with ALMA it not possible to observe different bands at the same time. For that reason we show a separate plot for every ALMA band. We see that the detection rate increases from band 3 to band 6 and, quite as expected, with observation time. From band 6 to band 7 the detection rates decrease again, because the high resolution of band 7 leads to a large number of resolved discs and thus to a decrease of flux per beam. The exact detection rate varies significantly from one hypothesis to another. For the hypotheses with $R=\mathrm{const}$, the detection rates for all but hypothesis 1 are very small. For the hypotheses with $R\propto L_{*}^{0.19}$ (7 to 10) the detection rates look a bit more promising.

We reiterate here that the sample we have chosen to work with in this paper is a sample of the closest M-stars. In order to avoid resolving discs to an extent that they are no longer detectable we could consider a more distant population of M-stars since the angular diameter of the disc is inversely proportional to the distance. Unfortunately, since the flux density is inversely proportional to the distance squared, the flux density per beam will still reduce with distance, thus showing that a more distant population of M star discs would also not benefit from ALMA observations.

\begin{table}
	\centering
	\caption{The resolution of each band and the fraction of discs that would be spatially resolved for two different assumptions about the disc radii.
}
	\label{tab:resolved_table}
	\begin{tabular}{lcccl}
		\hline
		 & resolution & R=const & R$\propto L_{*}^{0.19}$ \\
		\hline
		  Band 3 & $2.99''$ 	& 55.5\% & 20.4\% \\
		  Band 6 & $1.47''$ 	& 81.9\% & 40.6\% \\
		  Band 7 & $0.97''$ 	& 90.6\% & 54.4\% \\
		\hline
	\end{tabular}
	\\[3mm]
\textbf{Note:}	\\
The angular resolution in Table~\ref{tab:resolved_table} was calculated with  $ \theta_{res} = 0.574$ $\lambda$/$L_{80}$ \citep{ALMA-2019} with $\lambda$ being the observing wavelength and $L_{80}$ the 80th percentile of the ($u,v$) distance.
\end{table}

\subsubsection{Extragalactic confusion}

In the same way as for IRAM (subsection \ref{subsec:IRAMgalaxies}) we extrapolated the number of galaxies from band 6 to band 3 and 7. This was done for a 60~min ALMA observation assuming a >3$\sigma$ detection. There is a 0.8\%, 4.0\% and 5.6\% chance of a detectable galaxy within 1 beam of a star for band 3, 6 and 7, respectively. For band 7 the expected number of galaxies is higher than for bands 3 and 6. This could be a larger problem for observations. However, the number of resolved discs is also much higher for band 7 than for the other bands, which can help distinguish between a disc and a galaxy.

\subsubsection{Confusion due to stellar emission}
When considering unresolved observations we must also consider the contribution of the stellar flux density to the photometric flux density. Whilst the flux from the photosphere falls off with wavelength at millimetre wavelengths, similar to a blackbody, contributions from the chromosphere and corona can exceed the photosphere at such long wavelengths \citep[see e.g. ][]{liseau-et-al-2015}. This is particularly a problem for late type stars and can be hard to distinguish from disc emission. For instance, when AU Mic was first observed by ALMA, in addition to the clearly resolved debris disc, the central emission was found to be about six times higher than predicted for the photosphere. It initially was not clear whether this excess was emission from an asteroid belt or from the star \citep{macgregor-et-al-2013}, although further investigation clearly favours coronal emission \citep{cranmer-et-al-2013, daley-et-al-2019}. Similarly, \citet{anglada-et-al-2017} announced a tentative discovery of multiple dust rings around Proxima Centauri, however this star has long been known to be a flare star \citep{shapley-1951} and \citet{macgregor-et-al-2018} have since shown that the flares are responsible for the unresolved excess. Any sensitive surveys of M-stars would need to take this into consideration, although it is currently hard to quantify how large of an effect this would have since further research is necessary to better understand the flux distribution of M-stars at these wavelengths.

\subsubsection{Prospects for ALMA Observations}

Overall, our calculations show that a population of discs around M-stars similar to that around hotter stars could be detected with ALMA much more easily than with \emph{Herschel} when simply comparing the sensitivity of the observations with the expected total disc flux (Table~\ref{tab:ALMA_detection_prob}). However, the higher sensitivity of the observations does introduce potential issues due to extragalactic confusion (which is more severe for the shorter wavelengths)
and stellar confusion (which is more severe for the longer wavelengths). Resolving the discs would help alleviate these confusion issues, but also reduces the flux density per beam of the disc and thus the disc detectability. We also note that a limitation of our model is that it is reliant on the AFGK discs detected by \emph{Herschel}. As seen in Figure~\ref{fig:ALMA-shape}, ALMA extends the parameter space of detectable debris discs beyond what could be detected by \emph{Herschel} and so the detection rates reported in Table~\ref{tab:ALMA_detection_prob} should be considered lower limits. 

In conclusion, the high resolution of ALMA observations proves to be a hindrance to an unbiased survey aimed at detecting debris discs around M-stars.

\section{Discussion}

In view of these results, we argue that future observatories are necessary to help us find the missing population of debris discs around M-stars. One possibility would be to wait for far-infrared space missions of the next generation, such as \emph{Space Infrared Telescope for Cosmology and Astrophysics} \citep[\emph{SPICA};][]{roelfsema-et-al-2018} and \emph{Origins Space Telescope} \citep[\emph{OST};][]{ost-mission-2018}. Offering sensitivity by up to two orders of magnitude higher than \emph{Herschel}, these facilities would be able to probe the majority of the M-star debris discs predicted in the framework of our hypotheses, while also going much deeper in detecting debris discs of solar-type stars, catching the discs at least as tenuous as the dust disc in the Kuiper-belt region of our own Solar System. This would allow one to better determine how debris discs around M-stars relate to those around earlier type stars.
Another possibility would be to use large sub-mm single-dish telescopes like the proposed Atacama Large Aperture Submillimeter Telescope \citep[AtLAST;][]{holland-et-al-2019}. Being more sensitive than ALMA, this instrument would offer a resolution of $\approx 1.4''$ at $350\mum$, comparable to ALMA's in a compact configuration.

Observations with future observatories of this kind would not only promise a higher detection rate of discs around M-type stars. The chances to distinguish between different hypotheses would also be better.
Should detection rates be on the lower side, this would favour steeper dust mass -- stellar luminosity relations (hypotheses~5 or 10). Conversely, higher detection rates would be indicative of a weaker dependence of the dust mass on the stellar luminosity (hypotheses~1--2 or 6--7). Very high detection rates, above 20\% or so, would be suggestive of a population of large and cold debris discs around M-stars that do not have counterparts around earlier-type stars.

Obviously, the value of the future searches of M-star debris discs would extend far beyond pure statistics. Should they yield a set of new detections, and should some of the newly discovered discs of M-stars get resolved, this would allow one to infer disc radii and dust masses. This, in turn, would shed light onto the physics of debris discs around cool stars, which is not well understood at present. Further, combined with information (even observational upper limits) on planets in the same systems, disc parameters would enable deeper insights into the architecture of planetary systems of low-mass stars and into the history of their formation. For instance, should the majority of the newly discovered M-star discs be more compact than those around earlier-type primaries, this might be consistent with the fact that many M-stars host close-in low-mass planets, but only a few more distant planets in the giant mass range. Indeed, \citet{gaidos-2017} suggested that the density profiles on protoplanetary discs of M-stars may be steeper than those of solar type stars. This was deduced from the differences in planet masses and planet distributions. Effectively this would lead to more compact planet-forming discs around M-stars. Since debris discs are successors of protoplanetary ones, this could also be inherited by M-star debris discs. This would then be in line with hypotheses~6 through 10 in our analysis.

Another aspect is the dust mass. Should debris dust masses inferred from the new detections be abnormally low, this might be an indication that either the dust production rate in the discs of M-stars is low or the dust removal mechanisms in these discs are more efficient than in their counterparts of solar-type stars.  The former could be attributed, e.g., to the lower stirring level of the M-star discs \citep{thebault-wu-2008}, which might be consistent with the lack of massive planets as stirrers~--- or with the absence of large (Pluto-sized) embedded planetesimals \citep{pawellek-krivov-2015}, as was specifically inferred for the AU~Mic disc from its low vertical thickness \citep{daley-et-al-2019}.
The latter can be indicative of the role of the stellar winds \citep[e.g.][]{plavchan-et-al-2005,schueppler-et-al-2015} or intensive sweep-up of dust by coronal mass ejections \citep{misconi-1993,misconi-pettera-1995,osten-et-al-2013}.
	
\section{Conclusions}

In an attempt to clarify the reasons for the paucity of debris discs around M-type stars, in this paper we check whether in reality these discs could be as common as those around earlier-type stars, but could have just escaped frequent detection because of the observational limitations of the instruments used to search for them. We assume that M-star discs are ``similar'' to those around AFGK-stars, by formulating several hypotheses of what exactly that ``similarity'' may mean. Specifically, we speculate that the discs may either preserve nearly the same sizes around host stars from A to M~--- or may become somewhat more compact from earlier towards later-type primaries. In the same style, we assume that the dust mass in the discs may be either independent of the spectral type or gently decrease from more luminous to less luminous hosts. For each of the hypotheses, we compute the expected detection rates and compare them with those actually reported. 

As a test of our method, we first predicted the detection rates of discs around K-type stars, based on the AFG-star disc observations done by the \emph{\emph{Herschel}} Open Time Key Program DEBRIS \citep{thureau-et-al-2014,sibthorpe-et-al-2018}. We found the predicted rates to be consistent, within statistical uncertainties, with those found by the DEBRIS program \citep{sibthorpe-et-al-2018}, except for two hypotheses thus allowing us to rule out hypotheses 4 and 5 from the K-star data alone. With this test, we have validated the procedure before applying it to the M-star discs.

We computed the expected detection rates for M-stars, using the AFGK-stars reported by the DEBRIS program, and compared them with actual detection statistics (2~detections out of 94~stars probed, or $2.1^{+4.5}_{-1.7}$ at a 95\% confidence level) of that survey (Lestrade et al., in prep.). We found essentially the same result. The rates (1.6\%--6.4\%) are consistent with nearly all of the hypotheses, and we were not able to favour any of them. We identified only one hypothesis out of 10 considered that should formally be rejected. This was the assumption that discs of M-stars are more compact than, but as dusty as, those of earlier-type hosts (hypothesis~6 in Table~\ref{tab:summary_table}). For all other hypotheses, the assumption that discs of M-stars are as frequent as those of AFGK ones, would be fully consistent with the low incidence rates found by DEBRIS. This is in agreement with the conclusions drawn by \citet{morey-lestrade-2014}.
Our method confirms their results with a completely different approach. 

We tested which of the current-day instruments could be able to detect those M-star debris discs that we calculated before. First, we considered the new NIKA-2 instrument on IRAM, which operates at 1~mm and 2~mm. We tested it for three different observing times: 15~min, 30~min and 60~min. For the 15~min observation we obtained a detection probability between 0.6\%--4.5\%, depending on which hypothesis is assumed. These detection rates are slightly lower than those we estimated for the \emph{Herschel}/PACS DEBRIS survey. The minimum and maximum detection rates increase for a 60~min observation to 1.0\%--6.5\%.
We determine that there is a 6.6\% and 1.8\% change of finding a galaxy within 1 beam of a star for IRAM 1 mm and 2 mm, respectively, which is similar to the probabilities of detecting a debris disc.
With these results we do not favour IRAM NIKA-2 for the detection of a larger number of M-star discs. Nevertheless, one must consider that IRAM would partly detect different discs than \emph{Herschel}/PACS did, see Fig.~\ref{fig:IRAM-shape}.

Second, we made a prediction for a possible observation with ALMA in bands~3, 6 and 7. Analogous to the IRAM prediction, we tested three different observing times: 15~min, 30~min and 60~min. The detection rate    in band~3 ranges from 2.1\%--10.5\% for the 15~min observation to 3.1\%--13 .4\% for the 60~min observation. For band 6 the detection rate ranges from 3.9\%--15.1\% for 15~min integration time to 5.3\%--17.1\% for the 60~min observation. For band 7 the detection rate for 15~min is 4.3\%--15.8\% and increases to 4.5\%--17.6\% for the 60~min observation. Therefore, when only considering the sensitivity, our results imply that ALMA observations have the potential to discover many debris discs around M-stars, particularly with observations in bands 6 and 7. Unfortunately, there are a number of caveats that need to be taken into account with ALMA observations. 

Firstly, we considered the number of discs that would be resolved, as this decreases the amount of flux per beam and so the detectability of the discs. For hypotheses 1 to 5, where $R$=const, the number of resolved discs for band 3, 6 and 7 are 55.5\%, 81.9\% and 90.6\%, respectively. So for all bands, the majority of the discs would be resolved. The mean number of beams that are necessary to cover the complete discs are in this case 6.7 beams for band 3, 13.7 beams for band 6 and 20.6 beams for band 7. 
For hypotheses 6 to 10, where $R \propto L_{*}^{0.19}$, the number of resolved discs for band 3, 6 and 7 are 20.4\%, 40.6\% and 54.4\%, respectively. With a mean of 2.4 beams (band 3), 5.0 beams (band 6) and 7.5 beams (band 7) to cover the complete discs.
Considering the exact number of beams necessary to completely cover each disc and the MRS, we get the following detection rates for ALMA: band~3 ranges from 0.9\%--5.2\% (15~min) to 1.7\%--8.6\% (60~min), band 6 from 1.3\%--7.3\% for (15~min) to 2.3\%--11.0\% (60~min) and band 7 from 1.0\%--6.3\% (15~min) to 1.8\%--10.0\% (60~min).

Another aspect is the expected probability of having contaminating galaxies in the beam. It increases from 0.8\% in band 3 through 4.0\% in band 6 to 5.6\% in band 7. Especially in the latter case, given a maximum predicted detection rate of 10.0\% in band 7, background galaxies
could lead to a number of false disc candidates.  Since for band 7 also the number of resolved discs is the highest, it could, however, be possible to identify those galaxies.

In summary, we find that current observatories and instrumentation are not able to help us answer the question of how debris discs around M-stars relate to those around earlier type stars. Future observatories are necessary to help us find the missing population of debris discs around M-stars such as the far-infrared space missions \emph{SPICA} and \emph{OST}
or large sub-mm single-dish telescopes like AtLAST. 
Future surveys with facilities like these will show whether debris discs around M-stars are rare or rather ubiquitous. They may also allow one to distinguish between several hypotheses made here, enabling valuable constraints on salient features and physics of M-star debris discs. Combined with the results of planet searches, this would lead to better understanding of the architecture and formation circumstances of planetary systems around M-type stars.

\section*{Acknowledgements}

We are grateful to Grant Kennedy, Luca Matr\`a, and Mark Wyatt for enlightening discussions. Useful comments by the anonymous reviewer that helped to improve the manuscript are very much appreciated. This research was supported by the {\em Deutsche Forschungsgemeinschaft} (DFG), grants Kr~2164/13-2, Kr~2164/14-2 and Kr~2164/15-2.

\section*{Data availability}

The data underlying this article will be shared on reasonable request to the corresponding author.



\bsp	
\label{lastpage}

\input paper.bbl

\end{document}